\documentclass[11pt]{article}
\usepackage{amssymb,amsmath,amsfonts}
\usepackage{graphicx}
\usepackage{graphics}
\usepackage{eepic,epsfig}

\@addtoreset{equation}{section}

\makeatother

\begin{document}

\title{Hadamard function and the vacuum currents in braneworlds \\
with compact dimensions: Two-branes geometry}
\author{ S. Bellucci$^{1}$\thanks{%
E-mail: bellucci@lnf.infn.it }, A. A. Saharian$^{2}$\thanks{%
E-mail: saharian@ysu.am }, V. Vardanyan$^{2}$\thanks{%
E-mail: vardanyanv@gmail.com } \vspace{0.3cm} \\
\textit{$^1$ INFN, Laboratori Nazionali di Frascati,}\\
\textit{Via Enrico Fermi 40,00044 Frascati, Italy} \vspace{0.3cm}\\
\textit{$^2$ Department of Physics, Yerevan State University,}\\
\textit{1 Alex Manoogian Street, 0025 Yerevan, Armenia }}
\maketitle

\begin{abstract}
We evaluate the Hadamard function and the vacuum expectation value (VEV) of
the current density for a charged scalar field in the region between two
co-dimension one branes on the background of locally AdS spacetime with an
arbitrary number of toroidally compactified spatial dimensions. Along
compact dimensions periodicity conditions are considered with general values
of the phases and on the branes Robin boundary conditions are imposed for
the field operator. In addition, we assume the presence of a constant gauge
field. The latter gives rise to Aharonov-Bohm type effect on the vacuum
currents. There exists a range in the space of the Robin coefficients for
separate branes where the vacuum state becomes unstable. Compared to the
case of the standard AdS bulk, in models with compact dimensions the
stability condition imposed on the parameters is less restrictive. The
current density has nonzero components along compact dimensions only. These
components are decomposed into the brane-free and brane-induced
contributions. Different representations are provided for the latter well
suited for the investigation of the near-brane, near-AdS boundary and
near-AdS horizon asymptotics. The component along a given compact dimension
is a periodic function of the gauge field flux, enclosed by that dimension,
with the period of the flux quantum. An important feature, that
distinguishes the current density from the expectation values of the field
squared and energy-momentum tensor, is its finiteness on the branes. In
particular, for Dirichlet boundary condition the current density vanishes on
the branes. We show that, depending on the constants in the boundary
conditions, the presence of the branes may either increase or decrease the
current density compared with that for the brane-free geometry. Applications
are given to the Randall--Sundrum 2-brane model with extra compact
dimensions. In particular, we estimate the effects of the hidden brane on
the current density on the visible brane.
\end{abstract}

\bigskip


\bigskip

\section{Introduction}

\label{sec:introd}

In quantum field theory the vacuum is defined as the state of a quantum
field with zero number of quanta. The field operator does not commute with
the operator of the number of quanta and, hence, in the vacuum state the
field has no definite value. The corresponding quantum fluctuations are
known as zero-point or vacuum fluctuations. The properties of these
fluctuations and, hence, of the vacuum state, crucially depend on the
geometry of the background spacetime (for general reviews see \cite{Birr82}%
). Not surprisingly, exact results for the physical characteristics of the
vacuum can be found for highly symmetric backgrounds only. Continuing our
previous research \cite{Beze15,Bell15}, in this paper we investigate the
changes in the properties of the vacuum state for a charged scalar field
induced by three types of sources: by the curved geometry, by nontrivial
topology and by boundaries.

As a background geometry we will consider locally anti-de Sitter (AdS)
spacetime. AdS spacetime is the maximally symmetric solution of the vacuum
Einstein equations with a negative cosmological constant and because of its
high symmetry numerous physical problems are exactly solvable in this
geometry. In particular, quantum field theory in AdS background has long
been an active field of research. There are a number of reasons for that.
Much of the early interest in the seventies was motivated by principal
questions of the quantization procedure on curved backgrounds. Among the new
features, having no analogues in quantum field theory on the Minkowski bulk,
are the lack of global hyperbolicity and the presence of both regular and
irregular modes. In addition, the natural length scale of the AdS geometry
provides a convenient infrared regulator in interacting quantum field
theories without reducing the number of symmetries \cite{Call90}. The
natural appearance of AdS spacetime as a ground state in supergravity and
Kaluza-Klein theories and also as the near horizon geometry of the extremal
black holes and domain walls has triggered a further increase of interest to
quantum field theories on AdS bulk. This motivated the developement of a
parallel line of research, i.e. that of supersymmetric field theory models
in AdS background spacetime, see e.g. \cite{Bell86}.

The AdS geometry plays the crucial role in two recent developments in
high-energy physics such as the AdS/CFT correspondence and the braneworld
scenario. The AdS/CFT correspondence (see, for instance, \cite{Ahar00})
relates string theories or supergravity in the AdS bulk with a conformal
field theory localized on its boundary. This duality has many interesting
consequences and provides a powerful tool for the investigation of gauge
field theories in the strong coupling regime. Among the recent developments
of the AdS/CFT correspondence is the application to strong-coupling problems
in condensed matter physics. The braneworld scenario (for reviews see \cite%
{Ruba01}) offers a new perspective for the solution of the hierarchy problem
between the Planck and electroweak mass scales. The main idea to resolve the
large hierarchy is that the small coupling of four-dimensional gravity is
generated by the large physical volume of extra dimensions. Braneworlds
naturally appear in the string/M-theory context and present intriguing
possibilities to solve or to address from a different point of view various
problems in particle physics and cosmology.

The global geometry considered in the present paper will be different from
the standard AdS one. Namely, we will assume that a part of spatial
dimensions, described in Poincar\'{e} coordinates, are compactified to a
torus. Note that the extra compact dimensions are an inherent feature of
braneworld models arising from string and M-theories. The nontrivial
topology of the background space can have important physical implications in
quantum field theory. The periodicity conditions imposed on fields along
compact dimensions modify the spectrum for zero-point fluctuations and,
related to this, the vacuum expectation values (VEVs) of physical
observables are changed. A well-known effect of this kind, demonstrating the
relation between quantum phenomena and global properties of spacetime, is
the topological Casimir effect \cite{Eliz94}. The Casimir energy of bulk
fields induces a nontrivial potential for the compactification radius,
providing a stabilization mechanism for moduli fields and effective gauge
couplings. The Casimir effect has also been considered as an origin for the
dark energy in Kaluza-Klein-type and braneworld models \cite{Eliz01}.

For charged fields an important characteristic of the vacuum state is the
expectation value of the current density. In addition to describing the
local physical structure of the quantum field, the current acts as the
source in the Maxwell equations and plays an important role in modeling a
self-consistent dynamics involving the electromagnetic field. The VEV of the
current density for a charged scalar field in the background of locally AdS
spacetime with an arbitrary number of toroidally compactified spatial
dimensions has been considered in \cite{Beze15} (for a recent review of
quantum filed-theoretical effects in toroidal topology see \cite{Khan14}).
Both the zero and finite temperature expectation values of the current
density for charged scalar and fermionic fields in background of the flat
spacetime with toroidal dimensions were investigated in \cite{Beze13c,Bell10}%
. The vacuum current densities for charged scalar and Dirac spinor fields in
de Sitter spacetime with compact spatial dimensions are considered in \cite%
{Bell13b}. The effects of nontrivial topology induced by the
compactification of a cosmic string along its axis have been discussed in
\cite{Beze13}.

As the third source for the vacuum polarization, we will consider two
co-dimension one branes parallel to the AdS boundary. The effects induced by
a single brane were studied in \cite{Bell15}. The influence of boundaries on
the vacuum currents in topologically nontrivial flat spaces are studied in
\cite{Bell13,Bell15b} for scalar and fermionic fields. Note that, motivated
by the problems of radion stabilization and the cosmological constant
generation, the investigations of the vacuum energy and related forces for
branes on AdS bulk have attracted a great deal of attention (see, for
instance, the references in \cite{Eliz13}). The Casimir effect in
higher-dimensional generalizations of the AdS spacetime with compact
internal spaces has been discussed in \cite{Flac03,Saha06,Eliz07}.

The organization of the paper is as follows. The next section is devoted to
the description of the background geometry, the configuration of the branes,
the boundary conditions, and the field content. In section \ref{sec:Had}, we
evaluate the Hadamard function in the region between the branes. The single
brane contributions are explicitly separated and an integral representation
for the interference part is obtained well adapted for the investigation of
the VEVs for physical quantities bilinear in the field operator. In section %
\ref{sec:Curr}, the expression for the Hadamard function is used for the
investigation of the vacuum current in the region between the branes. The
behavior of the current density in various asymptotic regions of the
parameters is discussed. Numerical examples are presented in the case when
the Robin coefficients on separate branes are the same. The applications of
the results to the Randall-Sundrum 2-brane model with extra compact
dimensions are given in section \ref{sec:RS}. The main results of the paper
are summarized in section \ref{sec:Conc}. Alternative representations for
the Hadamard functions, adapted for the investigation of the near-brane
asymptotic of the vacuum current, are provided in Appendix.

\section{Field content, bulk and boundary geometries}

\label{sec:Geom}

First we will describe the bulk geometry. The corresponding metric tensor is
given by the $(D+1)$-dimensional line element
\begin{equation}
ds^{2}=g_{\mu \nu }dx^{\mu }dx^{\nu }=e^{-2y/a}\eta _{ik}dx^{i}dx^{k}-dy^{2},
\label{metric}
\end{equation}%
where $i,k=0,\ldots ,D-1$, $a$ is a constant, and $\eta _{ik}=\mathrm{diag}%
(1,-1,\ldots ,-1)$ is the metric tensor for $D$-dimensional Minkowski
spacetime. In addition to the $y$ coordinate, $-\infty <y\,+\infty $, we
will use the coordinate $z$, defined as $z=ae^{y/a}$, $0\leqslant z<\infty $%
. In terms of the latter, the line element is written in a manifestly
conformally flat form:%
\begin{equation}
ds^{2}=(a/z)^{2}(\eta _{ik}dx^{i}dx^{k}-dz^{2}).  \label{metric2}
\end{equation}%
The local geometry given by (\ref{metric2}) coincides with that for AdS
spacetime described in Poincar\'{e} coordinates. The hypersurfaces $z=0$ and
$z=\infty $ present the AdS boundary and horizon, respectively. The constant
$a$ is related to the Ricci scalar by the formula $R=-D(D+1)/a^{2}$ and the
metric tensor corresponding to (\ref{metric2}) is a solution of the vacuum
Einstein equations with a negative cosmological constant $\Lambda
=-D(D-1)a^{-2}/2$.

\begin{figure}[tbph]
\begin{center}
\epsfig{figure=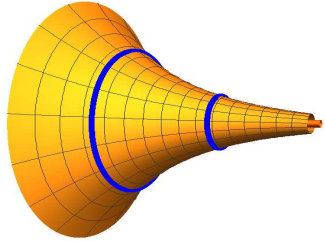,width=5cm,height=4cm}
\end{center}
\caption{The spatial section of the geometry at hand for $D=2$ embedded into
a 3-dimensional Euclidean space. The thick circles present the locations of
the branes.}
\label{fig1}
\end{figure}

The global properties of the geometry we are going to consider here will be
different from that for AdS spacetime. We assume that the subspace normal to
the $y$-coordinate has the topology $R^{p}\times T^{q}$, with $p$ and $q$
being integers such that $p+q=D-1$, and $T^{q}$ stands for a $q$-dimensional
torus. So, for the ranges of the coordinates $x^{i}$ in (\ref{metric}) one
has%
\begin{eqnarray}
-\infty &<&x^{i}<+\infty ,\;i=1,2,\ldots ,p,  \notag \\
0 &\leqslant &x^{i}\leqslant L_{i},\;i=p+1,\ldots ,D-1,  \label{Coord}
\end{eqnarray}%
with $L_{i}$ being the coordinate length of the $i$th compact dimension.
Note that the proper length measured by an observer with a fixed $z$ is
given by $L_{(p)i}=(a/z)L_{i}=e^{-y/a}L_{i}$. The latter decreases with
increasing $y$. This feature is seen in figure \ref{fig1} where we have
displayed the spatial geometry in the case $D=2$, embedded into the
3-dimensional Euclidean space. The circles correspond to the compact
dimension and the thick circles are the locations of the branes (see below).

Here we are interested in the VEV of the current density for a charged
scalar field $\varphi (x)$ in the background geometry specified above. In
addition, we will assume the presence of an external classical gauge field $%
A_{\mu }$. The dynamics of the field is governed by the equation
\begin{equation}
\left( g^{\mu \nu }D_{\mu }D_{\nu }+m^{2}+\xi R\right) \varphi (x)=0,
\label{fieldeq}
\end{equation}%
where $\xi $ is the curvature coupling parameter, $D_{\mu }=\nabla _{\mu
}+ieA_{\mu }$, with $\nabla _{\mu }$ being the covariant derivative
operator, $m$ and $e$ are the mass and the charge of the field quanta. The
most important special cases correspond to minimally and conformally coupled
fields with $\xi =0$ and $\xi =(D-1)/(4D)$, respectively. The background
topology is nontrivial and for the complete formulation of the problem, in
addition to the field equation, the periodicity conditions should be
specified along compact dimensions. Here we consider quasiperiodicity
conditions%
\begin{equation}
\varphi (t,x^{1},\ldots ,x^{l}+L_{l},\ldots ,y)=e^{i\alpha _{l}}\varphi
(t,x^{1},\ldots ,x^{l},\ldots ,y),  \label{PerC}
\end{equation}%
with constant phases $\alpha _{l}$, $l=p+1,\ldots ,D-1$ . As special cases,
these conditions include untwisted ($\alpha _{l}=0$) and twisted ($\alpha
_{l}=\pi $) scalars.

Now we turn to the description of the boundary geometry. It consists of two
co-dimension one branes, located at $y=y_{1}$ and $y=y_{2}$, $y_{1}<y_{2}$,
on which the field operator obeys the gauge invariant Robin boundary
conditions
\begin{equation}
(1+\beta _{j}n_{j}^{\mu }D_{\mu })\varphi (x)=0,\quad y=y_{j},  \label{Rob}
\end{equation}%
with $j=1,2$. Here, $\beta _{1}$ and $\beta _{2}$ are constants, $n_{j}^{\mu
}$ is the inward pointing (with respect to the region under consideration)
normal to the brane at $y=y_{j}$. In the region between the branes, $%
y_{1}\leqslant y\leqslant y_{2}$, in the coordinates $(x^{i},y)$ one has $%
n_{j}^{\mu }=\delta _{j}\delta _{D}^{\mu }$, where $\delta _{1}=1$ and $%
\delta _{2}=-1$. The locations of the branes in terms of the conformal
coordinate $z$ we will denote by $z_{1}$ and $z_{2}$, $z_{j}=ae^{y_{j}/a}$.
For the proper distance between the branes one has $y_{2}-y_{1}=a\ln
(z_{2}/z_{1})$. Boundary conditions of the type (\ref{Rob}) appear in a
number of physical problems, including the considerations of vacuum effects
for a confined charged scalar field in external fields \cite{Ambj83}, gauge
field theories, quantum gravity and supergravity \cite{Espo97,Luck91}, and
in models where the boundaries separate different gravitational backgrounds
\cite{Bell14b}. The Robin boundary conditions naturally arise in braneworld
models (see below). A more general class of boundary conditions in the
context of AdS/CFT correspondence, that include tangential derivatives of
the field on the boundary, has been discussed in \cite{Andr12,Andr12b}. In
the corresponding approach the boundary conditions are implemented by adding
the surface term in the action for a scalar field that contains a boundary
kinetic term. The latter leads to the modification of the standard
Klein-Gordon inner product by a boundary term. It has been shown that the
appropriate choice of the surface action makes the modes with non-Dirichlet
boundary conditions on the AdS boundary normalizable. However, because of
the lack of a manifestly positive inner product, ghosts may appear in the
bulk theory. The presence of the brane sufficiently far from the AdS
boundary can serve as a mechanism to banish these ghosts \cite{Andr12b}.

In what follows we will consider the gauge field configuration with constant
$A_{\mu }$. In this case, by the gauge transformation $\varphi
(x)=e^{-ie\chi (x)}\varphi ^{\prime }(x)$, $A_{\mu }=A_{\mu }^{\prime
}+\partial _{\mu }\chi (x)$, with the function $\chi (x)=A_{\mu }x^{\mu }$,
we can pass to a new gauge with the zero vector potential, $A_{\mu }^{\prime
}=0$. However, the vector potential of the former gauge will not completely
disappear from the problem. It will enter in the periodicity conditions for
the new field operator:%
\begin{equation}
\varphi ^{\prime }(t,x^{1},\ldots ,x^{l}+L_{l},\ldots ,y)=e^{i\tilde{\alpha}%
_{l}}\varphi ^{\prime }(t,x^{1},\ldots ,x^{l},\ldots ,y),  \label{PerC2}
\end{equation}%
where%
\begin{equation}
\tilde{\alpha}_{l}=\alpha _{l}+eA_{l}L_{l},\;l=p+1,\ldots ,D-1.  \label{alfl}
\end{equation}%
Hence, the presence of a constant gauge field is equivalent to the shift in
the phases of the quasiperiodicity conditions along compact dimensions. In
particular, nontrivial phases are generated for untwisted and twisted
scalars. The phase shift is expressed in terms of the magnetic flux $\Phi
_{l}$ enclosed by the $l$th compact dimension: $eA_{l}L_{l}=-2\pi \Phi
_{l}/\Phi _{0}$, with $\Phi _{0}=2\pi /e$ being the flux quantum (for
physical effects of gauge field fluxes in higher dimensional models with
compact dimensions see, for example, \cite{Doug07}).

\section{Hadamard function}

\label{sec:Had}

We are considering a free field theory (the only interactions are with the
background gravitational and electromagnetic fields) and all the information
on the properties of the quantum vacuum is encoded in two-point functions.
As such we will choose the Hadamard function, defined as the VEV $%
G(x,x^{\prime })=\langle 0|\varphi (x)\varphi ^{+}(x^{\prime })+\varphi
^{+}(x^{\prime })\varphi (x)|0\rangle $, where $|0\rangle $ corresponds to
the vacuum state. In what follows it will be assumed that the field is
prepared in the Poincar\'{e} vacuum state. The latter is realized by the
mode functions of the field which are obtained by solving the field equation
in Poincar\'{e} coordinates corresponding to (\ref{metric}) or (\ref{metric2}%
). The VEVs of physical observables, bilinear in the field operator, such as
the energy-momentum tensor and current density, are obtained from the
Hadamard function after some differentiations and limiting transition to the
coincidence limit of the arguments (with an appropriate renormalization). In
what follows we will present the evaluation procedure in the gauge with the
fields $(\varphi ^{\prime }(x),A_{\mu }^{\prime }=0)$, omitting the prime.

By expanding the field operator in terms of a complete set of positive- and
negative-energy mode functions $\{\varphi _{\sigma }^{(\pm )}(x)\}$ (upper
and lower signs respectively), specified by the set of quantum numbers $%
\sigma $ and obeying the quasiperiodicity and boundary conditions of the
problem at hand, the Hadamard function can be presented as the mode sum%
\begin{equation}
G(x,x^{\prime })=\sum_{\sigma }\sum_{s=\pm }\varphi _{\sigma
}^{(s)}(x)\varphi _{\sigma }^{(s)\ast }(x^{\prime }),  \label{Gmode}
\end{equation}%
where $\sum_{\sigma }$ includes the summation over the discrete quantum
numbers and the integration over the continuous ones. The problem under
consideration is plane symmetric and the mode function can be expressed in
the factorized form
\begin{equation}
\varphi _{\sigma }^{(\pm )}(x)=z^{D/2}Z_{\nu }(\lambda z)e^{ik_{r}x^{r}\mp
i\omega t},  \label{Modes}
\end{equation}%
where $k_{r}x^{r}=\sum_{r=1}^{D-1}k_{r}x^{r}$,
\begin{equation}
\omega =\sqrt{\lambda ^{2}+k^{2}},\;k^{2}=\sum_{l=1}^{D-1}k_{l}^{2}.
\label{lamb}
\end{equation}%
In (\ref{Modes}), $Z_{\nu }(x)$ is a cylinder function and%
\begin{equation}
\nu =\sqrt{D^{2}/4-D(D+1)\xi +m^{2}a^{2}}.  \label{nu}
\end{equation}%
For a conformally coupled massless scalar field $\nu =1/2$ and the problem
under consideration is conformally related to the problem in Minkowski
spacetime with two Robin boundaries (with the appropriate transformations of
the Robin coefficients, see below and also \cite{Saha03} for a general
plane-symmetric conformally flat bulk). In the case of imaginary values of $%
\nu $ the vacuum state becomes unstable \cite{Brei82}. In the discussion
below we assume the values of the parameters for which $\nu \geqslant 0$
(Breitenlohner--Freedman bound). In particular, this is the case for
minimally and conformally coupled fields.

For the components of the momentum along uncompact dimensions we have $%
-\infty <k_{l}<+\infty $, $l=1,\ldots ,p$, and the eigenvalues for the
components along compact dimensions are obtained from the conditions (\ref%
{PerC2}):
\begin{equation}
k_{l}=(2\pi n_{l}+\tilde{\alpha}_{l})/L_{l},\;l=p+1,\ldots ,D-1,  \label{kl}
\end{equation}%
where $n_{l}=0,\pm 1,\pm 2,\ldots $. In what follows, we will denote by $%
k_{(q)}^{2}$ the squared momentum in the compact subspace:%
\begin{equation}
k_{(q)}^{2}=\sum_{l=p+1}^{D-1}k_{l}^{2}=\sum_{l=p+1}^{D-1}(2\pi n_{l}+\tilde{%
\alpha}_{l})^{2}/L_{l}^{2}.  \label{kq}
\end{equation}%
Assuming that $|\tilde{\alpha}_{i}|\leqslant \pi $, for the lowest value of
this momentum, denoted here by $k_{(q)}^{(0)}$, one has
\begin{equation}
k_{(q)}^{(0)2}=\sum_{i=p+1}^{D-1}\tilde{\alpha}_{i}^{2}/L_{i}^{2}.
\label{kq0}
\end{equation}%
In particular, for an untwisted scalar field and for the zero gauge field
one has $k_{(q)}^{(0)}=0$.

The branes divide the space into three regions: $-\infty <y<y_{1}$, $%
y_{1}<y<y_{2}$, and $y>y_{2}$. In general, the curvature radius $a$ can be
different in these three sections, as the branes may separate different
phases of theory. In the braneworld scenario with two branes based on the
orbifolded version of the model the region between the branes is employed
only (see below). The Hadamard functions in the regions $y<y_{1}$ and $%
y>y_{2}$ coincide with the corresponding functions in the geometry of a
single brane located at $y=y_{1}$ and $y=y_{2}$, respectively, with the same
boundary conditions. The latter geometry is considered in \cite{Bell15} and
here we will be mainly concentrated on the region between the branes, $%
y_{1}\leqslant y\leqslant y_{2}$. In this region, the function $Z_{\nu
}(\lambda z)$ in (\ref{Modes}) is a linear combination of the Bessel and
Neumann functions $J_{\nu }(\lambda z)$ and $Y_{\nu }(\lambda z)$. Imposing
the boundary condition (\ref{Rob}) (with $D_{\mu }=\partial _{\mu }$) on the
brane $y=y_{1}$ we find%
\begin{equation}
Z_{\nu }(\lambda z)=C_{\sigma }g_{\nu }(\lambda z_{1},\lambda z),  \label{Z}
\end{equation}%
with the function%
\begin{equation}
g_{\nu }(u,v)=J_{\nu }(v)\bar{Y}_{\nu }^{(1)}(u)-\bar{J}_{\nu
}^{(1)}(u)Y_{\nu }(v).  \label{gnu}
\end{equation}%
Here and below, for a given function $F(x)$, the notations with the bars are
defined as
\begin{equation}
\bar{F}^{(j)}(x)=B_{j}xF^{\prime }(x)+A_{j}F(x),\quad j=1,2,  \label{Fbar}
\end{equation}%
where the coefficients are given by
\begin{equation}
B_{j}=\delta _{j}\beta _{j}/a,\;A_{j}=1+DB_{j}/2.  \label{A0}
\end{equation}%
Note that in the special cases $A_{j}=\pm \nu B_{j}$ one has
\begin{equation*}
\bar{F}_{\nu }^{(j)}(x)=\pm B_{j}xF_{\nu \mp 1}(x),
\end{equation*}%
for $F_{\nu }=J_{\nu }$ and $F_{\nu }=Y_{\nu }$. These special cases
correspond to the values
\begin{equation}
\beta _{j}/a=-\frac{\delta _{j}}{D/2\mp \nu },  \label{betSp}
\end{equation}%
for the Robin coefficients and, hence, $B_{j}=-1/(D/2\mp \nu )$.

From the boundary condition on the brane $y=y_{2}$ it follows that the
eigenvalues of $\lambda $ are solutions of the equation
\begin{equation}
\bar{J}_{\nu }^{(1)}(\lambda z_{1})\bar{Y}_{\nu }^{(2)}(\lambda z_{2})-\bar{Y%
}_{\nu }^{(1)}(\lambda z_{1})\bar{J}_{\nu }^{(2)}(\lambda z_{2})=0.
\label{Eigval}
\end{equation}%
Firstly we will assume that all the roots of this equation are real. The
changes in the evaluation procedure in the case when purely imaginary
eigenvalues are present for $\lambda $ will be discussed below. We denote by
$\lambda =\lambda _{n}$, $\lambda _{n}<\lambda _{n+1}$, $n=1,2,\ldots $, the
positive roots of (\ref{Eigval}). Note that, for a fixed interbrane distance
$y_{2}-y_{1}$ and Robin coefficients $\beta _{j}$, the product $z_{1}\lambda
_{n}$ does not depend on the location of the branes and on the lengths of
compact dimensions. The set of quantum numbers $\sigma $ specifying the mode
functions are given by $\sigma =(n,\mathbf{k}_{p},\mathbf{n}_{q})$, where $%
\mathbf{k}_{p}=(k_{1},\ldots ,k_{p})$ is the momentum in the non-compact
subspace and $\mathbf{n}_{q}=(n_{p+1},\ldots ,n_{D-1})$ determines the
momentum in the compact subspace. The normalization coefficient $C_{\sigma }$
in (\ref{Z}) is found from the condition
\begin{equation}
\int d^{D}x\,\sqrt{|g|}g^{00}\varphi _{\sigma }^{(s)}(x)\varphi _{\sigma
^{\prime }}^{(s^{\prime })\ast }(x)=\frac{\delta _{ss^{\prime }}}{2\omega }%
\delta _{nn^{\prime }}\delta (\mathbf{k}_{p}-\mathbf{k}_{p}^{\prime })\delta
_{\mathbf{n}_{q},\mathbf{n}_{q}^{\prime }},  \label{Norm}
\end{equation}%
where $s,s^{\prime }=+,-$ and the $y$-integration goes over the region
between the branes, $y_{1}\leqslant y\leqslant y_{2}$. By taking into
account that the function $g_{\nu }(\lambda z_{1},\lambda z)$ is a cylinder
function of the order $\nu $ with respect to the second argument (containing
the integration variable) and using the standard integral for the square of
the cylinder functions (see, for instance, \cite{Prud86}) we get the
following result%
\begin{equation}
|C_{\sigma }|^{2}=\frac{\pi ^{2}\lambda _{n}T_{\nu }(\chi ,z_{1}\lambda _{n})%
}{4\omega a^{D-1}\left( 2\pi \right) ^{p}V_{q}z_{1}},\quad \chi =\frac{z_{2}%
}{z_{1}},  \label{Csig}
\end{equation}%
where we have introduced the notation
\begin{equation}
T_{\nu }(\chi ,u)=u\left\{ \frac{\bar{J}_{\nu }^{(1)2}(u)}{\bar{J}_{\nu
}^{(2)2}(\chi u)}\left[ (\chi ^{2}u^{2}-\nu ^{2})B_{2}^{2}+A_{2}^{2}\right]
-(u^{2}-\nu ^{2})B_{1}^{2}-A_{1}^{2}\right\} ^{-1}.  \label{Tnu}
\end{equation}%
In (\ref{Csig}), $V_{q}=L_{p+1}\cdots L_{D-1}$ is the volume of the compact
subspace.

Substituting the mode functions into (\ref{Gmode}), the Hadamard function is
presented in the form%
\begin{eqnarray}
G(x,x^{\prime }) &=&\frac{a^{1-D}(zz^{\prime })^{D/2}}{2^{p+1}\pi
^{p-2}V_{q}z_{1}}\sum_{\mathbf{n}_{q}}\int d\mathbf{k}_{p}\,e^{ik_{r}\Delta
x^{r}}\sum_{n=1}^{\infty }\frac{\lambda _{n}}{\omega _{n}}  \notag \\
&&\times T_{\nu }(\chi ,z_{1}\lambda _{n})g_{\nu }(\lambda _{n}z_{1},\lambda
_{n}z)g_{\nu }(\lambda _{n}z_{1},\lambda _{n}z^{\prime })\cos (\omega
_{n}\Delta t),  \label{G}
\end{eqnarray}%
where $\Delta x^{r}=x^{r}-x^{\prime r}$, $\Delta t=t-t^{\prime }$, and $%
\omega _{n}=\sqrt{\lambda _{n}^{2}+k^{2}}$. The eigenvalues $\lambda _{n}$
are given implicitly, as roots of (\ref{Eigval}), and for that reason this
representation is not well adapted for the evaluation of the VEVs. Another
drawback is that the terms in the series with large $n$ are highly
oscillatory. A more convenient representation, free of these disadvantages,
is obtained by making use of the generalized Abel-Plana formula \cite%
{Saha87,Saha01}%
\begin{eqnarray}
\sum_{n=1}^{\infty }h(z_{1}\lambda _{n})T_{\nu }(\chi ,z_{1}\lambda _{n}) &=&%
\frac{2}{\pi ^{2}}\int_{0}^{\infty }\frac{h(x)dx}{\bar{J}_{\nu }^{(1)2}(x)+%
\bar{Y}_{\nu }^{(1)2}(x)}  \notag \\
&-&\frac{1}{2\pi }\int_{0}^{\infty }dx\,\Omega _{1\nu }(x,\chi x)\left[
h(ix)+h(-ix)\right] ,  \label{SumAbel}
\end{eqnarray}%
with the notations
\begin{equation}
\Omega _{1\nu }(u,v)=\frac{\bar{K}_{\nu }^{(2)}(v)}{\bar{K}_{\nu
}^{(1)}(u)F(u,v)},  \label{Om1}
\end{equation}%
and%
\begin{equation}
F(u,v)=\bar{K}_{\nu }^{(1)}(u)\bar{I}_{\nu }^{(2)}(v)-\bar{K}_{\nu }^{(2)}(v)%
\bar{I}_{\nu }^{(1)}(u).  \label{Fuv}
\end{equation}%
Here, $I_{\nu }(u)$ and $K_{\nu }(u)$ are the modified Bessel functions and
for the functions with the bars we use the notation defined by (\ref{Fbar}).
In the case of the function $h(x)$ corresponding to (\ref{G}), the
conditions of the validity for (\ref{SumAbel}) are satisfied if $z+z^{\prime
}+|\Delta t|<2z_{2}$. Note that in the coincidence limit and in the region
between the branes this condition is satisfied for points away from the
brane at $z=z_{2}$.

Let us denote by $G_{1}^{(1)}(x,x^{\prime })$ the contribution to the
Hadamard function coming from the first term in the right hand-side of (\ref%
{SumAbel}):%
\begin{eqnarray}
G_{1}^{(1)}(x,x^{\prime }) &=&\frac{\left( zz^{\prime }\right) ^{D/2}}{%
\left( 2\pi \right) ^{p}a^{D-1}V_{q}}\sum_{\mathbf{n}_{q}}\int d\mathbf{k}%
_{p}\,e^{ik_{r}\Delta x^{r}}\int_{0}^{\infty }d\lambda \,\lambda  \notag \\
&&\times \frac{\cos (\Delta t\sqrt{\lambda ^{2}+k^{2}})}{\sqrt{\lambda
^{2}+k^{2}}}\frac{g_{\nu }(\lambda z_{1},\lambda z)g_{\nu }(\lambda
z_{1},\lambda z^{\prime })}{\bar{J}_{\nu }^{(1)2}(\lambda z_{1})+\bar{Y}%
_{\nu }^{(1)2}(\lambda z_{1})}.  \label{Gsing}
\end{eqnarray}%
It coincides with the Hadamard function in the region $z>z_{1}$ in the
geometry of a single brane at $z=z_{1}$ and has been investigated in \cite%
{Bell15}. As a result, the application of (\ref{SumAbel}) leads to the
representation%
\begin{eqnarray}
G(x,x^{\prime }) &=&G_{1}^{(1)}(x,x^{\prime })-\frac{4(zz^{\prime })^{D/2}}{%
\left( 2\pi \right) ^{p+1}a^{D-1}V_{q}}\sum_{\mathbf{n}_{q}}\int d\mathbf{k}%
_{p}\,e^{ik_{r}\Delta x^{r}}\int_{k}^{\infty }du\,u  \notag \\
&&\times \frac{\Omega _{1\nu }(uz_{1},uz_{2})}{\sqrt{u^{2}-k^{2}}}X_{\nu
}^{(1)}(uz_{1},uz)X_{\nu }^{(1)}(uz_{1},uz^{\prime })\cosh (\Delta t\sqrt{%
u^{2}-k^{2}}),  \label{G1}
\end{eqnarray}%
where
\begin{equation}
X_{\nu }^{(j)}(u,v)=I_{\nu }(v)\bar{K}_{\nu }^{(j)}(u)-\bar{I}_{\nu
}^{(j)}(u)K_{\nu }(v),\;j=1,2.  \label{Gnuj}
\end{equation}%
For special values (\ref{betSp}) of the Robin coefficients one has%
\begin{equation}
\bar{I}_{\nu }^{(j)}(x)=B_{j}xI_{\nu \mp 1}(x),\;\bar{K}_{\nu
}^{(j)}(x)=-B_{j}xK_{\nu \mp 1}(x),  \label{BarSp}
\end{equation}%
with $B_{j}=-1/(D/2\mp \nu )$. The second term in the right-hand side of (%
\ref{G1}) is induced by the presence of the brane at $z=z_{2}$. Note that,
extracting the Hadamard function for the bulk in the absence of the branes, $%
G_{0}(x,x^{\prime })$, the function (\ref{Gsing}) is expressed as \cite%
{Bell15}
\begin{eqnarray}
G_{1}^{(1)}(x,x^{\prime }) &=&G_{0}(x,x^{\prime })-\frac{4\left( zz^{\prime
}\right) ^{D/2}}{\left( 2\pi \right) ^{p+1}a^{D-1}V_{q}}\sum_{\mathbf{n}%
_{q}}\int d\mathbf{k}_{p}\,e^{ik_{r}\Delta x^{r}}\,\int_{k}^{\infty }du\,
\notag \\
&&\times u\frac{\cosh (\Delta t\sqrt{u^{2}-k^{2}})}{\sqrt{u^{2}-k^{2}}}\frac{%
\bar{I}_{\nu }^{(1)}(uz_{1})}{\bar{K}_{\nu }^{(1)}(uz_{1})}K_{\nu
}(uz)K_{\nu }(uz^{\prime }),  \label{GsingDec}
\end{eqnarray}%
with the last term being the brane-induced contribution.

Another representation for the Hadamard function is obtained by using the
identity
\begin{eqnarray}
&&\frac{\bar{K}_{\nu }^{(2)}(uz_{2})}{\bar{I}_{\nu }^{(2)}(uz_{2})}I_{\nu
}(uz)I_{\nu }(uz^{\prime })-\frac{\bar{I}_{\nu }^{(1)}(uz_{1})}{\bar{K}_{\nu
}^{(1)}(uz_{1})}K_{\nu }(uz)K_{\nu }(uz^{\prime })  \notag \\
&&\qquad =\sum_{j=1,2}\delta _{j}\Omega _{j\nu }(uz_{1},uz_{2})X_{\nu
}^{(j)}(uz_{j},uz)X_{\nu }^{(j)}(uz_{j},uz^{\prime }),  \label{IdentIK}
\end{eqnarray}%
where
\begin{equation}
\Omega _{2\nu }(u,v)=\frac{\bar{I}_{\nu }^{(1)}(u)}{\bar{I}_{\nu
}^{(2)}(v)F(u,v)}.  \label{Om2}
\end{equation}%
Combining this with the expressions (\ref{G1}) and (\ref{GsingDec}), one
gets
\begin{eqnarray}
G(x,x^{\prime }) &=&G_{1}^{(2)}(x,x^{\prime })-\frac{a^{1-D}(zz^{\prime
})^{D/2}}{2^{p-1}\pi ^{p+1}V_{q}}\sum_{\mathbf{n}_{q}}\int d\mathbf{k}%
_{p}\,e^{ik_{r}\Delta x^{r}}\int_{k}^{\infty }du\,u  \notag \\
&&\times \frac{\Omega _{2\nu }(uz_{1},uz_{2})}{\sqrt{u^{2}-k^{2}}}X_{\nu
}^{(2)}(uz_{2},uz)X_{\nu }^{(2)}(uz_{2},uz^{\prime })\cosh (\Delta t\sqrt{%
u^{2}-k^{2}}).  \label{G2}
\end{eqnarray}%
In this formula, the function
\begin{eqnarray}
G_{1}^{(2)}(x,x^{\prime }) &=&G_{0}(x,x^{\prime })-\frac{a^{1-D}(zz^{\prime
})^{D/2}}{2^{p-1}\pi ^{p+1}V_{q}}\sum_{\mathbf{n}_{q}}\int d\mathbf{k}%
_{p}\,e^{ik_{r}\Delta x^{r}}\int_{k}^{\infty }du  \notag \\
&&\times u\frac{\bar{K}_{\nu }^{(2)}(uz_{2})}{\bar{I}_{\nu }^{(2)}(uz_{2})}%
\frac{I_{\nu }(uz)I_{\nu }(uz^{\prime })}{\sqrt{u^{2}-k^{2}}}\cosh (\sqrt{%
u^{2}-k^{2}}\Delta t)  \label{G12}
\end{eqnarray}%
is the Hadamard function in the geometry of a single brane at $y=y_{2}$ when
the brane $y=y_{1}$ is absent (see also \cite{Bell15}).

In the discussion above we have assumed that all the roots $\lambda $ of the
equation (\ref{Eigval}) are real. However, depending on the values of the
coefficients in the Robin boundary conditions on the branes, this equation
can have purely imaginary roots, $\lambda =i\eta $, $\eta >0$ (for some
special cases see below). For the corresponding modes the mode functions are
given by the expression%
\begin{equation}
\varphi _{(\mathrm{im})\sigma }^{(\pm )}(x)=C_{\sigma }^{(\mathrm{im}%
)}z^{D/2}X_{\nu }^{(1)}(\eta z_{1},\eta z)e^{ik_{r}x^{r}\mp i\omega (\eta
)t},  \label{phiim}
\end{equation}%
where $\omega (\eta )=\sqrt{k^{2}-\eta ^{2}}$ and the function $X_{\nu
}^{(1)}(\eta z_{1},\eta z)$ is defined by (\ref{Gnuj}). If $\eta
>k_{(q)}^{(0)}$, then for the modes with $k_{(q)}^{(0)}\leqslant k<\eta $
the energy is purely imaginary and the vacuum state becomes unstable. In
order to escape this instability, we will assume that%
\begin{equation}
\eta <k_{(q)}^{(0)}.  \label{Stabil}
\end{equation}%
Note that in the absence of compact dimensions any imaginary root for the
eigenvalue equation would lead to the vacuum instability. Hence, in models
with compact dimensions the constraints given by the stability condition are
less restrictive. The functions (\ref{phiim}) obey the boundary condition on
the brane at $y=y_{1}$. From the boundary condition on the second brane it
follows that $\eta $ is the root of the equation
\begin{equation}
F(\eta z_{1},\eta z_{2})=0,  \label{ImmodeEq}
\end{equation}%
with the function $F(u,v)$ defined by the expression (\ref{Fuv}). Of course,
this equation could directly be obtained from (\ref{Eigval}).

By using the integration formula
\begin{equation}
\int_{\eta z_{1}}^{\eta z_{2}}du\,uX_{\nu }^{(1)2}(\eta z_{1},u)=\frac{1}{2}%
\left[ (u^{2}+\nu ^{2})X_{\nu }^{(1)2}(\eta z_{1},u)-u^{2}(\partial
_{u}X_{\nu }^{(1)}(\eta z_{1},u))^{2}\right] _{\eta z_{1}}^{\eta z_{2}},
\label{IntG1}
\end{equation}%
from the normalization condition (\ref{Norm}) (with the replacement $\delta
_{nn^{\prime }}\rightarrow \delta _{\eta \eta ^{\prime }}$) for the
coefficient in (\ref{phiim}) we find the expression
\begin{equation}
|C_{\sigma }^{(\mathrm{im})}|^{2}=\frac{(2\pi )^{-p}a^{1-D}\eta ^{2}}{%
V_{q}\omega (\eta )\bar{I}_{\nu }^{(1)2}(\eta z_{1})}\left[ \sum_{j=1,2}%
\frac{A_{j}^{2}-(\eta ^{2}z_{j}^{2}+\nu ^{2})B_{j}^{2}}{\delta _{j}\bar{I}%
_{\nu }^{(j)2}(\eta z_{j})}\right] ^{-1}.  \label{Cim}
\end{equation}%
Here we have used the relations%
\begin{eqnarray}
X_{\nu }^{(1)}(\eta z_{1},\eta z_{1}) &=&-B_{1},  \notag \\
\lbrack \partial _{x}X_{\nu }^{(1)}(\eta z_{1},x)]_{x=\eta z_{1}}
&=&A_{1}/(\eta z_{1}).  \label{G11}
\end{eqnarray}%
and
\begin{eqnarray}
X_{\nu }^{(1)}(\eta z_{1},\eta z_{2}) &=&-B_{2}\frac{\bar{I}_{\nu
}^{(1)}(\eta z_{1})}{\bar{I}_{\nu }^{(2)}(\eta z_{2})},  \notag \\
\lbrack \partial _{x}X_{\nu }^{(1)}(\eta z_{1},x)]_{x=\eta z_{2}} &=&\frac{%
A_{2}}{\eta z_{2}}\frac{\bar{I}_{\nu }^{(1)}(\eta z_{1})}{\bar{I}_{\nu
}^{(2)}(\eta z_{2})}.  \label{G112}
\end{eqnarray}%
From (\ref{ImmodeEq}) it follows that $\bar{I}_{\nu }^{(1)}(\eta z_{1})/\bar{%
I}_{\nu }^{(2)}(\eta z_{2})=\bar{K}_{\nu }^{(1)}(\eta z_{1})/\bar{K}_{\nu
}^{(2)}(\eta z_{2})$ and, hence, in (\ref{Cim}) we can replace the $I$%
-functions by the $K$-functions.

By taking into account the equation (\ref{ImmodeEq}), it can be seen that%
\begin{equation}
\sum_{j=1,2}\frac{B_{j}^{2}(\eta ^{2}z_{j}^{2}+\nu ^{2})-A_{j}^{2}}{\delta
_{j}\bar{I}_{\nu }^{(j)2}(\eta z_{j})}=\frac{\eta \partial _{u}\left[
F(uz_{1},uz_{2})\right] _{u=\eta }}{\bar{I}_{\nu }^{(1)}(\eta z_{1})\bar{I}%
_{\nu }^{(2)}(\eta z_{2})}.  \label{RelC}
\end{equation}%
Now, for the contribution of the modes (\ref{phiim}) to the Hadamard
function, by using the relation (\ref{RelC}), we find the following
expression%
\begin{eqnarray}
G_{(\mathrm{im})}(x,x^{\prime }) &=&-\frac{2(zz^{\prime })^{D/2}}{(2\pi
)^{p}V_{q}a^{D-1}}\sum_{\mathbf{n}_{q}}\int d\mathbf{k}_{p}\,e^{ik_{r}\Delta
x^{r}}\sum_{\eta }\frac{\eta }{\omega (\eta )}\frac{\overline{I}_{\nu
}^{(2)}(\eta z_{2})}{\overline{I}_{\nu }^{(1)}(\eta z_{1})}  \notag \\
&&\times \frac{\cos [\omega (\eta )\Delta t]}{[\partial
_{u}F(uz_{1},uz_{2})]_{u=\eta }}X_{\nu }^{(1)}(\eta z_{1},\eta z)X_{\nu
}^{(1)}(\eta z_{1},\eta z^{\prime }).  \label{Gim}
\end{eqnarray}%
The contribution to the Hadamard function from the modes with real $\lambda $
is still given by the expression (\ref{G}). However, in the evaluation
procedure by using the Abel-Plana formula differences arise compared to the
case in the absence of purely imaginary roots. In the presence of the
imaginary roots $\lambda =\pm i\eta $ the function used in the derivation of
the Abel-Plana summation formula (see \cite{Saha87,Saha01}) has poles on the
imaginary axis. These poles should be avoided by semicircles of small radius
in the right half-plane. The integrals over these semicircles lead to the
term%
\begin{equation}
-\frac{i}{2}\sum_{x=\eta z_{1}}\frac{\bar{K}_{\nu }^{(2)}(\chi x)}{\bar{K}%
_{\nu }^{(1)}(x)}\frac{h(ix)-h(-ix)}{\partial _{x}F(x,\chi x)},
\label{AddTerm}
\end{equation}%
which should be added to the right-hand side of (\ref{SumAbel}). In
addition, in the presence of the imaginary roots, the second integral in (%
\ref{SumAbel}) is understood in the sense of the principal value. Now, we
can see that, after the application of the generalized Abel-Plana summation
formula (with the additional term (\ref{AddTerm})) to the series over $n$ in
(\ref{G}), the part of the Hadamard function coming from the term (\ref%
{AddTerm}) is equal to $-G_{(\mathrm{im})}(x,x^{\prime })$. Hence, this part
cancels the contribution of the purely imaginary modes in the Hadamard
function. As a result, the expressions (\ref{G1}) and (\ref{G2}) remain
valid in the presence of the imaginary modes with $\eta <k_{(q)}^{(0)}$.

In addition to the modes discussed above, a mode may be present for which $%
\lambda =0$ and, hence, $\omega =k$. \ For this mode the function $Z_{\nu }$
in (\ref{Modes}) is a linear combination of $z^{\nu }$ and $z^{-\nu }$. The
relative coefficient in this combination is determined from the boundary
condition at $z=z_{1}$ and the mode functions are presented as%
\begin{equation}
\varphi _{(\mathrm{s})\sigma }^{(\pm )}(x)=C_{\sigma }^{(\mathrm{s}%
)}z^{D/2}[(z/z_{1})^{\nu }-b_{1}(z/z_{1})^{-\nu }]e^{ik_{r}x^{r}\mp ikt},
\label{SpMode}
\end{equation}%
with the notation%
\begin{equation}
b_{j}=\frac{1+\left( D/2+\nu \right) \delta _{j}\beta _{j}/a}{1+\left(
D/2-\nu \right) \delta _{j}\beta _{j}/a},\;j=1,2.  \label{bj}
\end{equation}%
Here we have assumed that $\beta _{j}/a\neq -\delta _{j}/(D/2\mp \nu )$.
From the boundary condition at $z=z_{2}$ it follows that%
\begin{equation}
b_{2}\left( z_{2}/z_{1}\right) ^{2\nu }=b_{1}.  \label{Mode0Eq}
\end{equation}%
For a given interbrane distance, the equation (\ref{Mode0Eq}) gives the
relation between the Robin coefficients. For Dirichlet boundary condition on
the branes this equation has no solutions. For Neumann boundary condition it
has a solution in the case $\nu =D/2$ only and the corresponding mode
function does not depend on $z$. In the Robin case, if the coefficients $%
\beta _{j}$ are the same for both the branes, $\beta _{1}=\beta _{2}=\beta $%
, the equation (\ref{Mode0Eq}) has no solutions for $a/\beta \leqslant \nu
-D/2$ and for these $\beta $ there are no modes with $\lambda =0$. Note that
for a minimally coupled field $\nu \geqslant D/2$.

The coefficient $C_{\sigma }^{(\mathrm{s})}$ in (\ref{SpMode}) is determined
from the normalization condition. As a result, the normalized mode functions
for the special mode with $\lambda =0$ are presented in the form%
\begin{equation}
\varphi _{(\mathrm{s})\sigma }^{(\pm )}(x)=\Omega (z)\varphi _{(\mathrm{M}%
)\sigma }^{(\pm )}(x),  \label{SpMode2}
\end{equation}%
where%
\begin{equation}
\varphi _{(\mathrm{M})\sigma }^{(\pm )}(x)=\frac{e^{ik_{r}x^{r}\mp ikt}}{%
\sqrt{2\left( 2\pi \right) ^{p}kV_{q}}},  \label{phiM}
\end{equation}%
are the mode functions for a massless scalar field in $D$-dimensional
Minkowski spacetime with the spatial topology $R^{p}\times T^{q}$ and%
\begin{eqnarray}
\Omega (z) &=&z^{D/2}\frac{(z/z_{1})^{\nu }-b_{1}(z/z_{1})^{-\nu }}{%
a^{(D-1)/2}z_{1}}\left[ \frac{b_{1}^{2}}{2\nu -2}+b_{1}\right.  \notag \\
&&\left. -\frac{1}{2\nu +2}-\frac{b_{1}}{b_{2}}\chi ^{2}\left( \frac{%
b_{2}^{2}}{2\nu -2}+b_{2}-\frac{1}{2\nu +2}\right) \right] ^{-1/2}.
\label{Om}
\end{eqnarray}

In the cases $\beta _{j}/a=-\delta _{j}/(D/2\mp \nu )$ the mode functions
for the special mode with $\lambda =0$ have the form (\ref{SpMode2}) with
the conformal function%
\begin{equation}
\Omega ^{2}(z)=\frac{2\left( 1\mp \nu \right) z^{D\mp 2\nu }}{%
a^{D-1}(z_{2}^{2\mp 2\nu }-z_{1}^{2\mp 2\nu })}.  \label{Omsp}
\end{equation}%
In the case $\nu =D/2$ and for Neumann boundary condition, for the mode with
$\lambda =0$ one has $\varphi _{(0)\sigma }^{(\pm )}(x)=C_{\sigma }^{(%
\mathrm{s})}e^{ik_{r}x^{r}\mp ikt}$ and for the corresponding function $%
\Omega (z)$ one gets
\begin{equation}
\Omega ^{2}(z)=\frac{\left( D-2\right) z_{1}^{D-2}}{%
a^{D-1}[1-(z_{1}/z_{2})^{D-2}]}.  \label{OmN}
\end{equation}%
This case will be considered in the numerical examples below.

As a consequence of the relation (\ref{SpMode2}), the contribution of the
mode with $\lambda =0$ to the Hadamard function, $G_{(\mathrm{s}%
)}(x,x^{\prime })$, is expressed in terms of the corresponding function for
a massless scalar field in $D$-dimensional Minkowski spacetime with the
spatial topology $R^{p}\times T^{q}$. Denoting the latter by $G_{R^{p}\times
T^{q}}^{(\mathrm{M})}(x,x^{\prime })$, one has $G_{(\mathrm{s})}(x,x^{\prime
})=\Omega (z)\Omega (z^{\prime })G_{R^{p}\times T^{q}}^{(\mathrm{M}%
)}(x,x^{\prime })$, or by making use of the expression for the Minkowskian
function:
\begin{equation}
G_{(\mathrm{s})}(x,x^{\prime })=\frac{2\Omega (z)\Omega (z^{\prime })}{%
\left( 2\pi \right) ^{p+1/2}V_{q}}\sum_{\mathbf{n}_{q}}e^{ik_{l}\Delta
x^{l}}k_{(q)}^{p-1}f_{(p-1)/2}(k_{(q)}\sqrt{\sum\nolimits_{l=1}^{p}(\Delta
x^{l})^{2}-\left( \Delta t\right) ^{2}}).  \label{GRs}
\end{equation}%
where $k_{l}\Delta x^{l}=\sum_{l=p+1}^{D-1}k_{l}\Delta x^{l}$, and%
\begin{equation}
f_{\mu }(x)=x^{-\mu }K_{\mu }(x).  \label{fmu}
\end{equation}%
Note that the dependence on the mass of the field in this expression appears
through the parameter $\nu $ in (\ref{Om}).

Under the condition (\ref{Mode0Eq}), the contribution (\ref{GRs}) coming
from the special mode with $\lambda =0$ should be added to the part (\ref{G}%
) for the modes with $\lambda =\lambda _{n}$. However, we can show that the
representations (\ref{G1}) and (\ref{G2}) are not changed. Indeed, in the
presence of the mode with $\lambda =0$ the function $h(x)$ in the
generalized Abel-Plana formula (\ref{SumAbel}) corresponding to the series
over $n$ in (\ref{G}) has a simple pole at $x=0$. Now, rotating the
integration contour in the derivation of the Abel-Plana formula we should
avoid this pole by arcs of a circle of small radius. The terms coming from
the integrals over these arcs exactly cancel the contribution (\ref{GRs}) of
the special mode. As a result, the formulas (\ref{G1}) and (\ref{G2}) remain
valid in the presence of the mode $\lambda =0$ as well.

\section{VEV of the current density}

\label{sec:Curr}

Having the Hadamard function, one can evaluate the VEVs of various local
physical observables bilinear in the field. The VEV of the energy-momentum
tensor for a scalar field in the geometry with two branes is investigated in
\cite{Knap04,Saha05,Shao10} for the background with trivial topology and in
\cite{Saha06} for models with extra compact subspaces. Our main interest
here is the VEV of the current density, $\langle 0|j_{\mu }(x)|0\rangle
\equiv \langle j_{\mu }(x)\rangle $. For a charged scalar field the
corresponding operator is given by the expression
\begin{equation}
j_{\mu }(x)=ie[\varphi ^{+}(x)D_{\mu }\varphi (x)-(D_{\mu }\varphi
^{+}(x))\varphi (x)].  \label{jmu}
\end{equation}%
The VEV is obtained from the Hadamard function by making use of the formula
\begin{equation}
\langle j_{\mu }(x)\rangle =\frac{i}{2}e\lim_{x^{\prime }\rightarrow
x}(\partial _{\mu }-\partial _{\mu }^{\prime }+2ieA_{\mu })G(x,x^{\prime }),
\label{jl1}
\end{equation}%
First of all, we can see that $\langle j_{\mu }\rangle =0$ for $\mu
=0,1,\ldots ,p,D$. Hence, the VEVs of the charge density and of the current
density components along uncompact dimensions (including the one
perpendicular to the branes) vanish.

By using the expressions (\ref{G1}) and (\ref{G2}) for the Hadamard function
and integrating over the momentum in the uncompact subspace, for the
component of the vacuum current along the $l$th compact dimension one finds
two equivalent representations%
\begin{eqnarray}
\langle j^{l}\rangle &=&\langle j^{l}\rangle _{0}+\langle j^{l}\rangle
_{1}^{(j)}-\frac{eC_{p}z^{D+2}}{2^{p-1}a^{D+1}V_{q}}\sum_{\mathbf{n}%
_{q}}k_{l}\,\int_{k_{(q)}}^{\infty }dx\,x  \notag \\
&&\times (x^{2}-k_{(q)}^{2})^{(p-1)/2}\Omega _{j\nu }(xz_{1},xz_{2})X_{\nu
}^{(j)2}(xz_{j},xz),  \label{jl2dec}
\end{eqnarray}%
with $j=1,2$ and $l=p+1,\ldots ,D-1$. Here we have introduced the notation%
\begin{equation}
C_{p}=\frac{\pi ^{-(p+1)/2}}{\Gamma ((p+1)/2)}.  \label{Cp}
\end{equation}%
In the formula (\ref{jl2dec}), $\left\langle j_{l}\right\rangle _{0}$ is the
current density in the absence of the branes and $\left\langle
j^{l}\right\rangle _{1}^{(j)}$ is the current density induced by the
presence of the brane at $y=y_{j}$ when the second brane is absent. Hence,
the last term in the right-hand side can be interpreted as the contribution
induced by the second brane at $y=y_{j^{\prime }}$, $j^{\prime }=1,2$, $%
j^{\prime }\neq j$.

The contribution $\langle j^{l}\rangle _{0}$ is investigated in \cite{Beze15}
and is given by the expression%
\begin{eqnarray}
\langle j^{l}\rangle _{0} &=&\frac{2ea^{-1-D}L_{l}}{(2\pi )^{(D+1)/2}}\sum_{%
\mathbf{n}_{q}}n_{l}\sin (\tilde{\alpha}_{l}n_{l})\cos (\sum_{i\neq l}\tilde{%
\alpha}_{i}n_{i})  \notag \\
&&\times q_{\nu -1/2}^{(D+1)/2}(1+\sum_{i}n_{i}^{2}L_{i}^{2}/(2z^{2})),
\label{jl0}
\end{eqnarray}%
where $q_{\alpha }^{\mu }(x)=(x^{2}-1)^{-\mu /2}e^{-i\pi \mu }Q_{\alpha
}^{\mu }(x)$, with $Q_{\alpha }^{\mu }(x)$ being the associated Legendre
function of the second kind. In the region $z_{1}\leqslant z\leqslant z_{2}$%
, for the single brane-induced parts in (\ref{jl2dec}) one has%
\begin{eqnarray}
\langle j^{l}\rangle _{1}^{(1)} &=&-\frac{eC_{p}z^{D+2}}{2^{p-1}a^{D+1}V_{q}}%
\sum_{\mathbf{n}_{q}}k_{l}\int_{k_{(q)}}^{\infty }dx\,x(x^{2}-k_{(q)}^{2})^{%
\frac{p-1}{2}}\frac{\bar{I}_{\nu }^{(1)}(xz_{1})}{\bar{K}_{\nu
}^{(1)}(xz_{1})}K_{\nu }^{2}(xz),  \notag \\
\langle j^{l}\rangle _{1}^{(2)} &=&-\frac{eC_{p}z^{D+2}}{2^{p-1}a^{D+1}V_{q}}%
\sum_{\mathbf{n}_{q}}k_{l}\int_{k_{(q)}}^{\infty }dx\,x(x^{2}-k_{(q)}^{2})^{%
\frac{p-1}{2}}\frac{\bar{K}_{\nu }^{(2)}(xz_{2})}{\bar{I}_{\nu
}^{(2)}(xz_{2})}I_{\nu }^{2}(xz).  \label{jl12}
\end{eqnarray}%
The properties of these single brane contributions are discussed in \cite%
{Bell15}.

All the contributions to the VEV of the current density and, hence, the
total current as well, are periodic functions of the phases $\tilde{\alpha}%
_{i}$ with the period equal $2\pi $. In particular, the current is a
periodic function of the magnetic flux enclosed by compact dimensions with
the period equal to the flux quantum. The VEV of the component for the
current density along the $l$th compact dimension is an odd function of the
phase $\tilde{\alpha}_{l}$ corresponding to the same direction and an even
function of the remaining phases $\tilde{\alpha}_{i}$, $i\neq l$. The
appearance of nonzero vacuum currents discussed here is a consequence of the
nontrivial spatial topology (though influenced by the local geometry and
boundaries). This is an Aharonov-Bohm type effect related to the sensitivity
of the wave function phase to the global geometry. For $\tilde{\alpha}%
_{i}=\pi m_{i}$, with $m_{i}$ being an integer, the current density
vanishes. The relation (\ref{alfl}) shows two interrelated reasons for the
appearance of the currents: nontrivial phases in the periodicity conditions
and the magnetic flux enclosed by compact dimensions.

By taking into account the expressions for the single brane-induced parts (%
\ref{jl12}), the total current density in the region between the branes can
also be presented in the form%
\begin{eqnarray}
\langle j^{l}\rangle &=&\langle j^{l}\rangle _{0}-\frac{eC_{p}z^{D+2}}{%
2^{p-1}a^{D+1}V_{q}}\sum_{\mathbf{n}_{q}}k_{l}\int_{k_{(q)}}^{\infty }dx\,x
\notag \\
&&\times (x^{2}-k_{(q)}^{2})^{\frac{p-1}{2}}\left[ \frac{\bar{K}_{\nu
}^{(1)}(xz_{1})\bar{I}_{\nu }^{(2)}(xz_{2})}{\bar{K}_{\nu }^{(2)}(xz_{2})%
\bar{I}_{\nu }^{(1)}(xz_{1})}-1\right] ^{-1}  \notag \\
&&\times \left[ I_{\nu }(xz)\frac{X_{\nu }^{(1)}(xz_{1},xz)}{\bar{I}_{\nu
}^{(1)}(xz_{1})}-K_{\nu }(xz)\frac{X_{\nu }^{(2)}(xz_{2},xz)}{\bar{K}_{\nu
}^{(2)}(xz_{2})}\right] .  \label{jl3}
\end{eqnarray}%
The second term in the right-hand side is the brane-induced contribution.
Alternatively, extracting the single-brane contributions we can write the
following decomposition%
\begin{equation}
\langle j^{l}\rangle =\langle j^{l}\rangle _{0}+\sum_{j=1,2}\langle
j^{l}\rangle _{1}^{(j)}+\langle j^{l}\rangle _{\mathrm{int}},
\label{jldecInt}
\end{equation}%
with the interference part%
\begin{eqnarray}
\langle j^{l}\rangle _{\mathrm{int}} &=&-\frac{eC_{p}z^{D+2}}{%
2^{p-1}a^{D+1}V_{q}}\sum_{\mathbf{n}_{q}}k_{l}\,\int_{k_{(q)}}^{\infty }dx\,x
\notag \\
&&\times (x^{2}-k_{(q)}^{2})^{\frac{p-1}{2}}\left[ \frac{\bar{K}_{\nu
}^{(1)}(xz_{1})\bar{I}_{\nu }^{(2)}(xz_{2})}{\bar{I}_{\nu }^{(1)}(xz_{1})%
\bar{K}_{\nu }^{(2)}(xz_{2})}-1\right] ^{-1}  \notag \\
&&\times \left[ I_{\nu }(xz)\frac{X_{\nu }^{(2)}(xz_{2},xz)}{\bar{I}_{\nu
}^{(2)}(xz_{2})}-K_{\nu }(xz)\frac{X_{\nu }^{(1)}(xz_{1},xz)}{\bar{K}_{\nu
}^{(1)}(xz_{1})}\right] .  \label{jlint}
\end{eqnarray}%
The integrand in this expression decays exponentially in the upper limit for
all points including those on the branes..

Let us consider some limiting cases of the general formulas. In the limit of
the large curvature radius, $a\gg y_{j},m^{-1}$, one has $z\approx a+y$, $%
z_{j}\approx a+y_{j}$, and both the order and arguments of the modified
Bessel functions in the integrand of (\ref{jl2dec}) are large. By using the
corresponding uniform asymptotic expansions (given, for example, in \cite%
{Abra72}), to the leading order, the result for the geometry of two parallel
Robin plates on the Minkowski bulk with the topology $R^{p+1}\times T^{q}$
(see \cite{Bell15b}) is obtained:%
\begin{eqnarray}
\langle j^{l}\rangle ^{(\mathrm{M})} &=&\langle j^{l}\rangle _{0}^{(\mathrm{M%
})}+\frac{eC_{p}}{2^{p}V_{q}}\sum_{\mathbf{n}_{q}}k_{l}\int_{\sqrt{%
m^{2}+k_{(q)}^{2}}}^{\infty }dx\,(x^{2}-m^{2}-k_{(q)}^{2})^{\frac{p-1}{2}}
\notag \\
&&\times \frac{2+\sum_{j=1,2}c_{j}(x)e^{2x|y-y_{j}|}}{%
c_{1}(x)c_{2}(x)e^{2x(y_{2}-y_{1})}-1}.  \label{jlM}
\end{eqnarray}%
where%
\begin{equation}
c_{j}(x)=\frac{\beta _{j}x-1}{\beta _{j}x+1},\;j=1,2.  \label{cj}
\end{equation}%
In (\ref{jlM}), the current density in the boundary-free Minkowskian
geometry with compact dimensions is given by the formula \cite{Beze13}%
\begin{equation}
\langle j^{l}\rangle _{0}^{(\mathrm{M})}=\frac{2eL_{l}m^{D+1}}{(2\pi
)^{(D+1)/2}}\sum_{\mathbf{n}_{q}}n_{l}\sin (n_{l}\tilde{\alpha}_{l})\cos
(\sum_{i\neq l}\tilde{\alpha}_{i}n_{i})f_{\frac{D+1}{2}}(m(%
\sum_{i}n_{i}^{2}L_{i}^{2})^{1/2}),  \label{jl0a}
\end{equation}%
with $f_{\mu }(x)$ defined by (\ref{fmu}).

For a conformally coupled massless field one has $\nu =1/2$ and the modified
Bessel functions are expressed in terms of the elementary functions. For the
current density in the region between the branes one finds the expression%
\begin{eqnarray}
\langle j^{l}\rangle &=&(z/a)^{D+1}\left[ \langle j^{l}\rangle _{0}^{\mathrm{%
(M)}}+\frac{eC_{p}}{2^{p}V_{q}}\sum_{\mathbf{n}_{q}}k_{l}\int_{k_{(q)}}^{%
\infty }dx\,\right.  \notag \\
&&\times \left. (x^{2}-k_{(q)}^{2})^{\frac{p-1}{2}}\frac{2+\sum_{j=1,2}%
\tilde{c}_{j}(xz_{j})e^{2x|z-z_{j}|}}{\tilde{c}_{1}(xz_{1})\tilde{c}%
_{2}(xz_{2})e^{2x(z_{2}-z_{1})}-1}\right] ,  \label{jlConf}
\end{eqnarray}%
where we have introduced the notation%
\begin{equation}
\tilde{c}_{j}(u)=\frac{u-a/\beta _{j}-\delta _{j}(D-1)/2}{u+a/\beta
_{j}+\delta _{j}(D-1)/2}.  \label{cjtilde}
\end{equation}%
Note that the expression in the square brackets of (\ref{jlConf}), with the
functions $\tilde{c}_{j}(u)$ replaced by $c_{j}(u)$ from (\ref{cj}),
coincides with the corresponding result in the region between two Robin
boundaries in Minkowski bulk with spatial topology $R^{p+1}\times T^{q}$.
The difference in the functions $\tilde{c}_{j}(u)$ and $c_{j}(u)$ is related
to that, though the bulk geometry is conformally flat and the field equation
is conformally invariant, the coefficient in the Robin boundary condition is
not conformally invariant.

As it has been shown in \cite{Bell15b} for the Minkowski bulk and in \cite%
{Bell15} for the geometry of a single brane on AdS bulk, unlike the VEVs of
the field squared and of the energy-momentum tensor, the current density is
finite on the branes. For the geometry under consideration, the VEV\ of the
current density on the brane is obtained from (\ref{jl2dec}) with $z=z_{j}$.
By taking into account that $X_{\nu }^{(j)}(xz_{j},xz_{j})=-B_{j}$, we get%
\begin{eqnarray}
\langle j^{l}\rangle _{z=z_{j}} &=&\langle j^{l}\rangle _{0}+\langle
j^{l}\rangle _{1,z=z_{j}}^{(j)}-\frac{eC_{p}z_{j}^{D+2}B_{j}^{2}}{%
2^{p-1}a^{D+1}V_{q}}\sum_{\mathbf{n}_{q}}k_{l}\,  \notag \\
&&\times \int_{k_{(q)}}^{\infty }dx\,x(x^{2}-k_{(q)}^{2})^{\frac{p-1}{2}%
}\Omega _{j\nu }(xz_{1},xz_{2}),  \label{jlonBr}
\end{eqnarray}%
where the last term is the current density induced by the second brane on
the brane at $z=z_{j}$. For Dirichlet boundary condition the single brane
and the second brane induced parts and, hence, also the total current,
vanish on the branes.

In the limit when the left brane tends to the AdS boundary, $%
z_{1}\rightarrow 0$, one gets%
\begin{eqnarray}
\langle j^{l}\rangle &\approx &\langle j^{l}\rangle _{0}+\langle
j^{l}\rangle _{1}^{(2)}-\frac{2^{2-p-2\nu }eC_{p}}{a^{D+1}V_{q}}\frac{%
A_{1}+B_{1}\nu }{A_{1}-B_{1}\nu }\frac{z_{1}^{2\nu }z^{D+2}}{\nu \Gamma
^{2}(\nu )}\sum_{\mathbf{n}_{q}}k_{l}\,  \notag \\
&&\times \int_{k_{(q)}}^{\infty }dx\,x^{2\nu +1}(x^{2}-k_{(q)}^{2})^{\frac{%
p-1}{2}}\frac{X_{\nu }^{(2)2}(xz_{2},xz)}{\bar{I}_{\nu }^{(2)2}(xz_{2})}.
\label{jlLim1}
\end{eqnarray}%
To the leading order we obtain the VEV in the geometry of a single brane at $%
z=z_{2}$. The contribution coming from the brane at $z=z_{1}$, corresponding
to the last term in (\ref{jlLim1}), decays as $z_{1}^{2\nu }$. When the
right brane is close to the AdS horizon, $z_{2}\rightarrow \infty $, our
starting point will be the expression (\ref{jl2dec}) with $j=1$. In the
limit under consideration one has%
\begin{equation}
\Omega _{1\nu }(xz_{1},xz_{2})\approx (2\delta _{B_{2}0}-1)\frac{\pi
e^{-2xz_{2}}}{\bar{K}_{\nu }^{(1)2}(xz_{1})}.  \label{Omas1}
\end{equation}%
The dominant contribution to the integral in the right-hand side of (\ref%
{jl2dec}) comes from the region near the lower limit of the integration and
from the term with $\mathbf{n}_{q}=0$ with the minimal value of $%
k_{(q)}=k_{(q)}^{(0)}$. In the leading order one gets%
\begin{eqnarray}
\langle j^{l}\rangle &\approx &\langle j^{l}\rangle _{0}+\langle
j^{l}\rangle _{1}^{(1)}+\frac{(1-2\delta _{B_{2}0})e\tilde{\alpha}_{l}z^{D+2}%
}{2^{p}\pi ^{(p-1)/2}a^{D+1}L_{l}V_{q}}  \notag \\
&&\times \,\frac{X_{\nu }^{(1)2}(k_{(q)}^{(0)}z_{1},k_{(q)}^{(0)}z)}{\bar{K}%
_{\nu }^{(1)2}(k_{(q)}^{(0)}z_{1})e^{2k_{(q)}^{(0)}z_{2}}}%
(k_{(q)}^{(0)}/z_{2})^{(p+1)/2},  \label{jlLim2}
\end{eqnarray}%
with $k_{(q)}^{(0)}z_{2}\gg 1$. Hence, when the right brane tends to the AdS
horizon, its contribution to the current density, for a fixed value of $z$,
is suppressed by the factor $\exp (-2k_{(q)}^{(0)}z_{2})$.

Now we turn to the asymptotics in the limiting cases for the lengths of
compact dimensions. Firstly, consider the limit when the length of the one
of compact dimensions, say $L_{r}$, $r\neq l$, is much larger than other
length scales, $L_{r}\gg L_{i},z_{1}$. In this case the dominant
contribution to the series over $n_{r}$ in (\ref{jl3}) comes from large
values of $|n_{r}|$ and the corresponding summation can be replaced by the
integration in accordance with $\sum_{n_{r}=-\infty }^{+\infty }\rightarrow
(L_{r}/\pi )\int_{0}^{\infty }dk_{r}$. Then, passing to a new integration
variable $u=\sqrt{x^{2}-k_{(q)}^{2}}$ and introducing polar coordinates in
the plane $(k_{r},u)$, after the integration over the angular part, we can
see that, to the leading order, the result is obtained for the current
density in the model where the $r$th dimension is uncompactified. In the
opposite limit of small lengths of the $r$th dimension, $L_{r}\ll L_{i},z_{1}
$, under the condition $|\tilde{\alpha}_{r}|<\pi $, in the expression (\ref%
{jl3}) the contribution of the term with $n_{r}=0$ dominates. The behavior
of the component of the current density along the $l$th compact dimension
crucially depends whether the phase $\tilde{\alpha}_{r}$ is zero or not. In
the case $\tilde{\alpha}_{r}=0$, the leading term obtained from the
right-hand side coincides with the current density in the $D$-dimensional
model, with the excluded $r$th compact dimension divided by $%
L_{(p)r}=aL_{r}/z$. Recall that the latter is the proper length of the $r$th
dimension measured by an observer with the fixed coordinate $z$. For the
case $\tilde{\alpha}_{r}\neq 0$, the arguments of the modified Bessel
functions in the integrand of (\ref{jl3}) are large. By making use of the
corresponding asymptotic formulas, we can see that the contribution of the
single brane at $z=z_{j}$ is suppressed by the factor $\exp (-2|\tilde{\alpha%
}_{r}||z-z_{j}|/L_{r})$, whereas the interference part decays as $\exp [-2|%
\tilde{\alpha}_{r}|(z_{2}-z_{1})/L_{r}]$.

If the length of the $l$th compact dimension is much smaller than the
remaining lengths, $L_{l}\ll L_{i}$, the main contribution to the series
over $\mathbf{n}_{q-1}=(n_{p+1},\ldots ,n_{l-1},n_{l+1},\ldots ,n_{D-1})$ in
(\ref{jl2dec}) comes from large values of $|n_{i}|$, $i=p+1,\ldots ,D-1$, $%
i\neq l$. In this case the corresponding summation can be replaced by the
integration in accordance with
\begin{equation}
\sum_{\mathbf{n}_{q-1}}f(k_{(q-1)})\rightarrow \frac{2^{2-q}\pi
^{-(q-1)/2}V_{q}}{L_{l}\Gamma ((q-1)/2)}\int_{0}^{\infty }du\,u^{q-2}f(u),
\label{SumtoInt}
\end{equation}%
where $k_{(q-1)}^{2}=k_{(q)}^{2}-k_{l}^{2}$. As the next step, instead of $x$
we introduce a new integration variable $w=\sqrt{x^{2}-u^{2}-k_{l}^{2}}$.
Passing to the polar coordinates in the plane $(u,w)$, after the evaluation
of the integral over the angular variable, we find
\begin{eqnarray}
\langle j^{l}\rangle  &\approx &\langle j^{l}\rangle _{0}+\langle
j^{l}\rangle _{1}^{(j)}-\frac{2^{3-D}\pi ^{(1-D)/2}ez^{D+2}}{%
a^{D+1}L_{l}\Gamma ((D-1)/2)}\sum_{n_{l}=-\infty }^{+\infty
}k_{l}\,\int_{|k_{l}|}^{\infty }dx\,x  \notag \\
&&\times (x^{2}-k_{l}^{2})^{\frac{D-3}{2}}\Omega _{j\nu
}(xz_{1},xz_{2})X_{\nu }^{(j)2}(xz_{j},xz).  \label{SmLl}
\end{eqnarray}%
A similar transformation is done with the single brane part $\langle
j^{l}\rangle _{1}^{(j)}$ and the right-hand side of (\ref{SmLl}) coincides
with the corresponding result in the model with a single compact dimension
of the length $L_{l}$.

When, in addition to the condition $L_{l}\ll L_{i}$, one has $L_{l}\ll z_{j}$%
, in the integration range of (\ref{SmLl}) the arguments of the modified
Bessel functions are large and we use the corresponding asymptotic
expressions \cite{Abra72}. For the single brane and interference parts one
gets%
\begin{eqnarray}
\langle j^{l}\rangle _{1}^{(j)} &\approx &-\gamma _{j}\frac{%
eL_{l}(z/L_{l})^{D+1}\mathrm{sgn}(\tilde{\alpha}_{l})}{(4\pi
)^{(D-1)/2}a^{D+1}}\frac{|\tilde{\alpha}_{l}|^{(D-1)/2}e^{-2|\tilde{\alpha}%
_{l}||z-z_{j}|/L_{l}}}{(|z-z_{j}|/L_{l})^{(D-1)/2}},  \notag \\
\langle j^{l}\rangle _{\mathrm{int}} &\approx &\gamma _{1}\gamma _{2}\frac{%
2eL_{l}(z/L_{l})^{D+1}\mathrm{sgn}(\tilde{\alpha}_{l})}{(4\pi
)^{(D-1)/2}a^{D+1}}\,\frac{|\tilde{\alpha}_{l}|^{(D-1)/2}e^{-2|\tilde{\alpha}%
_{l}|(z_{2}-z_{1})/L_{l}}}{[(z_{2}-z_{1})/L_{l}]^{(D-1)/2}},
\label{jlLargeL}
\end{eqnarray}%
with $L_{l}\ll |z-z_{j}|$. Here, $\gamma _{j}=2\delta _{0B_{j}}-1$ with $%
j=1,2$. As it is seen, both the single brane and interference contributions
decay exponentially and the decay of the interference part is stronger. In
the same limit, for the boundary-free part one has
\begin{equation}
\langle j^{l}\rangle _{0}\approx \frac{2e\Gamma ((D+1)/2)}{\pi
^{(D+1)/2}a^{D+1}}L_{l}(z/L_{l})^{D+1}\sum_{n=1}^{\infty }\frac{\sin (\tilde{%
\alpha}_{l}n)}{n^{D}},  \label{Llsm0}
\end{equation}%
and it dominates in the total VEV. Hence, in the limit under consideration
the brane-induced parts are mainly concentrated near the branes within the
range $|z-z_{j}|\lesssim L_{l}$.

Another representation for the VEV of the current density is obtained from
the decomposition (\ref{Gdecl1}) for the Hadamard function, where the part $%
G_{l}(x,x^{\prime })$, given by (\ref{Gl2}), is induced by the
compactification of the $l$th dimension. The first term in the right-hand
side does not contribute to the current density and, after the integrations,
in the absence of the modes with $\lambda ^{2}<0$, for the VEV we find the
following expression%
\begin{eqnarray}
\langle j^{l}\rangle &=&\frac{ea^{-1-D}z^{D+2}}{2(2\pi
)^{p/2-1}V_{q}L_{l}^{p}z_{1}}\sum_{s=1}^{\infty }\frac{\sin \left( s\tilde{%
\alpha}_{l}\right) }{s^{p+1}}\sum_{n=1}^{\infty }\lambda _{n}T_{\nu }(\chi
,\lambda _{n}z_{1})\,  \notag \\
&&\times g_{\nu }^{2}(\lambda _{n}z_{1},\lambda _{n}z)\sum_{\mathbf{n}%
_{q-1}}g_{p/2+1}(sL_{l}\sqrt{\lambda _{n}^{2}+k_{(q-1)}^{2}}),
\label{jlAlt0}
\end{eqnarray}%
with the function%
\begin{equation}
g_{\mu }(x)=x^{\mu }K_{\mu }(x).  \label{gmu}
\end{equation}%
In the model with a single compact dimension $x^{l}$, the corresponding
formula is obtained from (\ref{jlAlt0}) putting $p=D-2$, $k_{(q-1)}=0$, $%
V_{q}=L_{l}$ and omitting the summation $\sum_{\mathbf{n}_{q-1}}$. The
vanishing of the vacuum currents on the branes in the case of Dirichlet
boundary condition is explicitly seen from (\ref{jlAlt0}) by taking into
account that $g_{\nu }(\lambda _{n}z_{1},\lambda _{n}z_{j})=0$ for $j=1,2$.
Due to the presence of the Macdonald function, the series in the right-hand
side are strongly convergent for all values of $z$. In particular, the
representation (\ref{jlAlt0}) explicitly shows the finiteness of the current
density on the branes. For the VEV of the current density on the brane at $%
z=z_{j}$ from (\ref{jlAlt0}) one gets%
\begin{eqnarray}
\langle j^{l}\rangle _{z=z_{j}} &=&-\frac{8ea^{-1-D}B_{j}^{2}z_{j}^{D+2}}{%
(2\pi )^{p/2+1}V_{q}L_{l}^{p}}\sum_{s=1}^{\infty }\frac{\sin \left( s\tilde{%
\alpha}_{l}\right) }{s^{p+1}}\sum_{n=1}^{\infty }\frac{\lambda _{n}^{2}}{%
\bar{J}_{\nu }^{(j)2}(z_{j}\lambda _{n})}\,  \notag \\
&&\times \left[ \sum_{i=1,2}\frac{(z_{i}^{2}\lambda _{n}^{2}-\nu
^{2})B_{i}^{2}+A_{i}^{2}}{\delta _{i}\bar{J}_{\nu }^{(i)2}(z_{i}\lambda _{n})%
}\right] ^{-1}\sum_{\mathbf{n}_{q-1}}g_{p/2+1}(sL_{l}\sqrt{\lambda
_{n}^{2}+k_{(q-1)}^{2}}),  \label{jlzj}
\end{eqnarray}%
for $j=1,2$.

In the presence of the modes with $\lambda ^{2}\leqslant 0$ their
contribution to the current density should be separately added to the
right-hand side of (\ref{jlAlt0}). For the modes $\lambda =i\eta $, $\eta >0$%
, assuming that $\eta <k_{(q-1)}^{(0)}$, with $k_{(q-1)}^{(0)}$ being the
minimal value for $k_{(q-1)}$, the corresponding contribution to the current
density is formally obtained from (\ref{jlAlt0}) by making the replacements $%
\lambda _{n}\rightarrow i\eta $, $\sum_{n}\rightarrow \sum_{\eta }$. If $|%
\tilde{\alpha}_{i}|\leqslant \pi $, then one has $k_{(q-1)}^{(0)2}=%
\sum_{i=p+1,\neq l}^{D-1}\tilde{\alpha}_{i}^{2}/L_{i}^{2}$. For the possible
special mode with $\lambda =0$ its contribution to the current density is
given by the expression%
\begin{equation}
\langle j^{l}\rangle _{(\mathrm{s})}=\frac{4e\Omega ^{2}(z)(z/a)^{2}}{(2\pi
)^{p/2+1}V_{q}L_{l}^{p}}\sum_{n=1}^{\infty }\frac{\sin \left( n\tilde{\alpha}%
_{l}\right) }{n^{p+1}}\sum_{\mathbf{n}_{q-1}}g_{p/2+1}(nL_{l}k_{(q-1)}).
\label{jlSp}
\end{equation}%
The factor $(z/a)^{2}$ in this expression arises when one passes from the
covariant component of the current density to the contravariant one. In
models with a single compact dimension from here we get%
\begin{equation}
\langle j^{l}\rangle _{(\mathrm{s})}=\frac{2e\Gamma (D/2)\Omega
^{2}(z)(z/a)^{2}}{\pi ^{D/2}L^{D-1}}\sum_{n=1}^{\infty }\frac{\sin \left( n%
\tilde{\alpha}_{l}\right) }{n^{D-1}}.  \label{jlSp1}
\end{equation}%
In particular, for Neumann boundary conditions on both the branes and for $%
\nu =D/2$, the factor $\Omega ^{2}(z)$ does not depend on $z$ and is given
by (\ref{OmN}).

The representation (\ref{jlAlt0}) is well adapted for the investigation of
the asymptotic behavior for large values of $L_{l}$ compared with the other
length scales. In this limit one has $L_{l}\lambda _{n}\gg 1$ and the
argument of the function $g_{p/2+1}(x)$ is large. The dominant contribution
to the VEV comes from the mode with the lowest $\lambda _{n}$ and from the
term with $s=1$. By using the asymptotic for the function $g_{p/2+1}(x)$, it
is seen that the current density is suppressed by the factor $\exp (-L_{l}%
\sqrt{\lambda _{1}^{2}+k_{(q-1)}^{(0)2}})$. If the mode with $\lambda =0$ is
present, the corresponding asymptotic directly follows from (\ref{jlSp}).
For $k_{(q-1)}^{(0)}>0$, the contribution of the term with $n=1$ and $%
\mathbf{n}_{q-1}=0$ dominates and the corresponding current density decays
as $\exp (-L_{l}k_{(q-1)}^{(0)})$. This decay is weaker than for the modes
with positive $\lambda ^{2}$. In the case $k_{(q-1)}^{(0)}=0$, the decay of
the contribution of the zero mode goes down like a power-law, i.e. $%
1/L_{l}^{p+1}$.

The limit of small distances between the branes, compared with the AdS
curvature radius, corresponds to the ratio $z_{2}/z_{1}$ close to 1. With
decreasing $z_{2}/z_{1}$ the eigenvalues $\lambda _{n}$ increase and tend to
infinity in the limit $z_{2}/z_{1}\rightarrow 1$. With this feature, from (%
\ref{jlAlt0}) it follows that the contribution of the modes with positive $%
\lambda ^{2}$ to the VEV of the current density tends to zero in the limit
when the distance between the branes tends to 0. This is not necessary to be
the case for the contribution of the special mode with $\lambda =0$. For
example, in the case of Neumann boundary condition and for $\nu =D/2$, the
current density from the special mode is given by (\ref{jlSp}) with $\Omega
^{2}(z)$ defined by (\ref{OmN}). It diverges in the limit $%
z_{2}/z_{1}\rightarrow 1$.

An alternative representation for the current density in the region between
the branes is obtained by using the result (\ref{Gl3}):%
\begin{eqnarray}
\langle j_{l}\rangle &=&\langle j^{l}\rangle _{0}+\langle j^{l}\rangle
_{1}^{(1)}+\frac{4ea^{1-D}z^{D}}{(2\pi )^{p/2+1}V_{q}L_{l}^{p}}%
\sum_{n=1}^{\infty }\frac{\sin \left( n\tilde{\alpha}_{l}\right) }{n^{p+1}}%
\sum_{\mathbf{n}_{q-1}}\int_{k_{(q-1)}}^{\infty }dx\,  \notag \\
&&\times xw_{p/2+1}(nL_{l}\sqrt{x^{2}-k_{(q-1)}^{2}})\Omega _{1\nu
}(xz_{1},xz_{2})X_{\nu }^{(1)2}(xz_{1},xz).  \label{jlAlt}
\end{eqnarray}%
with the notation%
\begin{equation}
w_{\nu }(x)=x^{\nu }J_{\nu }(x).  \label{wnu}
\end{equation}%
By taking into account the corresponding formula from \cite{Bell15} for the
contribution $\langle j^{l}\rangle _{1}^{(1)}$, this expression is presented
as%
\begin{eqnarray}
\langle j_{l}\rangle &=&\langle j^{l}\rangle _{0}+\frac{4ea^{1-D}z^{D}}{%
(2\pi )^{p/2+1}V_{q}L_{l}^{p}}\sum_{n=1}^{\infty }\frac{\sin \left( n\tilde{%
\alpha}_{l}\right) }{n^{p+1}}\sum_{\mathbf{n}_{q-1}}\int_{k_{(q-1)}}^{\infty
}dx\,  \notag \\
&&\times xw_{p/2+1}(nL_{l}\sqrt{x^{2}-k_{(q-1)}^{2}})\left[ \frac{\bar{K}%
_{\nu }^{(1)}(xz_{1})\bar{I}_{\nu }^{(2)}(xz_{2})}{\bar{K}_{\nu
}^{(2)}(xz_{2})\bar{I}_{\nu }^{(1)}(xz_{1})}-1\right] ^{-1}  \notag \\
&&\times \left[ \frac{X_{\nu }^{(1)}(xz_{1},xz)}{\bar{I}_{\nu }^{(1)}(xz_{1})%
}I_{\nu }(xz)-\frac{X_{\nu }^{(2)}(xz_{2},xz)}{\bar{K}_{\nu }^{(2)}(xz_{2})}%
K_{\nu }(xz)\right] .  \label{jlAlt2}
\end{eqnarray}%
In the presence of the mode with $\lambda ^{2}<0$, the representations (\ref%
{jlAlt}) and (\ref{jlAlt2}) are valid under the condition $\eta
<k_{(q-1)}^{(0)}$.

\begin{figure}[tbph]
\begin{center}
\epsfig{figure=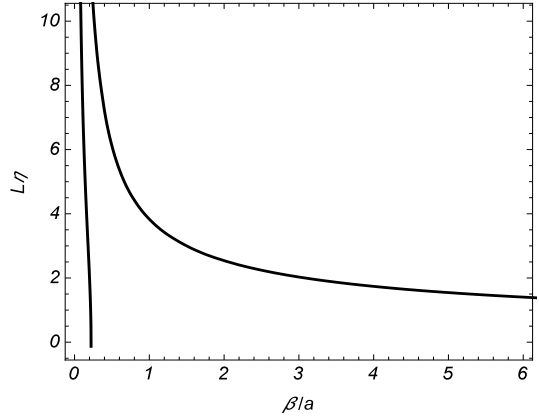,width=7.cm,height=5.5cm}
\end{center}
\caption{The imaginary modes for $\protect\lambda $ versus the Robin
coefficient for fixed values $z_{1}/L=0.5$, $z_{2}/L=1$, and $\tilde{\protect%
\alpha}=\protect\pi /2$.}
\label{fig2}
\end{figure}

In the figures below all the graphs are plotted for a minimally coupled
massless scalar field in $D=4$ (except the figure in section \ref{sec:RS},
where we consider the model with $D=5$) in the model with a single compact
dimension ($q=1$, $p=D-2$) of the length $L$ and with the phase $\tilde{%
\alpha}_{l}=\tilde{\alpha}$. For this model one has $\nu =2$ and, hence, $%
\nu =D/2$. As it has been discussed above, in this case for Neumann boundary
condition there is a special mode with $\lambda =0$. The corresponding
contribution to the current density is given by (\ref{jlSp1}) with $\Omega
^{2}(z)$ from (\ref{OmN}). In the numerical examples with Robin boundary
conditions we assume that the Robin coefficients for the branes are the
same: $\beta _{1}=\beta _{2}\equiv \beta $. In this case there are no modes
with imaginary $\lambda $ for $\beta \leqslant 0$. In order to see the
behavior of the imaginary modes as a function of the Robin coefficient in
the range $\beta >0$, in figure \ref{fig2} we have plotted the roots of the
equation (\ref{ImmodeEq}), multiplied by $L$, as a function of $\beta /a$
for fixed values $z_{1}/L=0.5$ and $z_{2}/L=1$. In the range $0<\beta
/a<0.2206$, for a given $\beta /a$ there are two roots. For $\beta /a>0.2206$
there is a single root which decreases with increasing $\beta /a$. For a
given $\tilde{\alpha}$, the vacuum is stable if $\beta \leqslant 0$ or $%
\beta >\beta _{c}$ where $\beta _{c}$ is the root of the equation $L\eta
(\beta /a)=\tilde{\alpha}$. This root is the abscissa of the intersection
point of the left curve in figure \ref{fig2} with the horizontal line $L\eta
=\tilde{\alpha}$. For example, in the case $\tilde{\alpha}=\pi /2$ (in the
numerical evaluations below we take this value of the phase) one has $\beta
_{c}/a\approx 4.845$.

\begin{figure}[tbph]
\begin{center}
\epsfig{figure=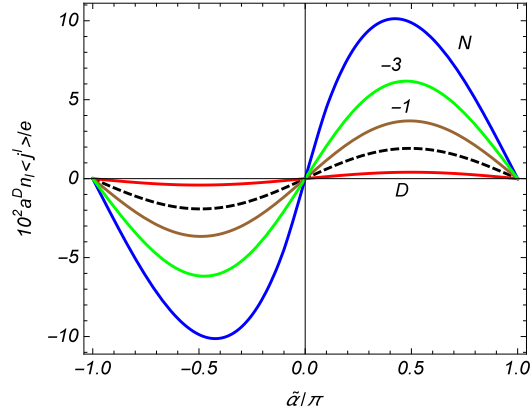,width=7.cm,height=5.5cm}
\end{center}
\caption{The vacuum current density in the region between the branes as a
function of the phase in the quasiperiodicity condition along the compact
dimension. The graphs are plotted for scalar fields with Dirichlet, Neumann
and Robin (for $\protect\beta /a=-1,-3$, numbers near the curves) boundary
conditions and for fixed values $z_{1}/L=0.5$, $z_{2}/L=1$, $z/L=0.75$. The
dashed curve corresponds to the current density in the absence of branes.}
\label{fig3}
\end{figure}

In the figures we plot the graphs for the charge flux density through the $%
(D-1)$-dimensional spatial hypersurface $x^{l}=\mathrm{const}$. The latter
is given by the quantity $n_{l}\langle j^{l}\rangle $, where $n_{l}=a/z$ is
the normal to the hypersurface. This quantity is the current density
measured by an observer with a fixed value for the coordinate $z$. Indeed,
in order to discuss the physics from the point of view of that observer, it
is convenient to introduce rescaled coordinates $x^{\prime i}=(a/z)x^{i}$.
With these coordinates the warp factor in the metric for the subspace
parallel to the branes is equal to one and they are physical coordinates of
the observer. For the current density in these coordinates one has $\langle
j^{\prime l}\rangle =(a/z)\langle j^{l}\rangle $ which is exactly the
quantity presented in the graphs below. In figure \ref{fig3} we have
displayed the dependence of the current density on the phase $\tilde{\alpha}$
for fixed values $z_{1}/L=0.5$, $z_{2}/L=1$, $z/L=0.75$. The graphs are
plotted for Dirichlet, Neumann and Robin (for $\beta /a=-1,-3$, numbers near
the curves) boundary conditions. The dashed curve presents the current
density in the geometry without branes. Recall that the current density is
an odd periodic function of $\tilde{\alpha}$ with the period $2\pi $.

\begin{figure}[tbph]
\begin{center}
\epsfig{figure=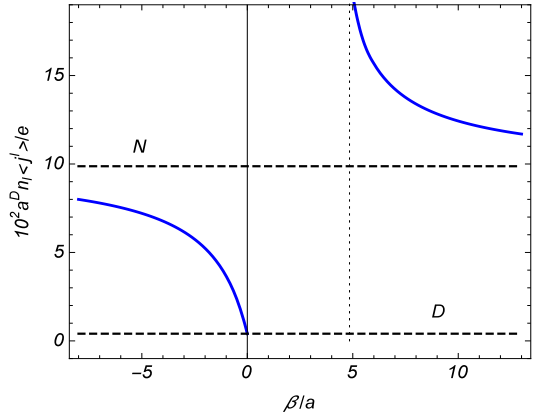,width=7.cm,height=5.5cm}
\end{center}
\caption{The VEV of the current density as a function of the Robin
coefficient for $z_{1}/L=0.5$, $z_{2}/L=1$, $z/L=0.75$, and $\tilde{\protect%
\alpha}=\protect\pi /2$. The dashed lines present the current densities for
Dirichlet and Neumann boundary conditions. }
\label{fig4}
\end{figure}

As it is seen from figure \ref{fig3}, for $\beta \leqslant 0$ the current
density is an increasing function of $|\beta |$. In order to show the
dependence of the current density on the coefficient in Robin boundary
condition, in figure \ref{fig4} the current density is plotted versus $\beta
/a$ for $z_{1}/L=0.5$, $z_{2}/L=1$, $z/L=0.75$, and $\tilde{\alpha}=\pi /2$.
As we have noted above, for these values of the parameters and in the region
$0<\beta /a<4.845$ there are modes with imaginary $\lambda =i\eta $ for
which $L\eta >\tilde{\alpha}$. This means that in this region the vacuum
state is unstable. In figure \ref{fig4}, the instability region is between
the ordinate axis and the dotted vertical line, corresponding to $\beta
/a=4.845$. The dashed horizontal lines correspond to the current density in
the cases of Ditichlet and Neumann boundary conditions. As we could expect,
for large values of $|\beta |$ the results for Robin boundary condition tend
to the one for Neumann condition, whereas for $\beta \rightarrow -0$ we
obtain the result for Dirichlet boundary condition. For $\beta <0$ ($\beta >0
$) the modulus of the current density for Robin boundary condition is
smaller (larger) than that for the Neumann case.

\begin{figure}[tbph]
\begin{center}
\begin{tabular}{cc}
\epsfig{figure=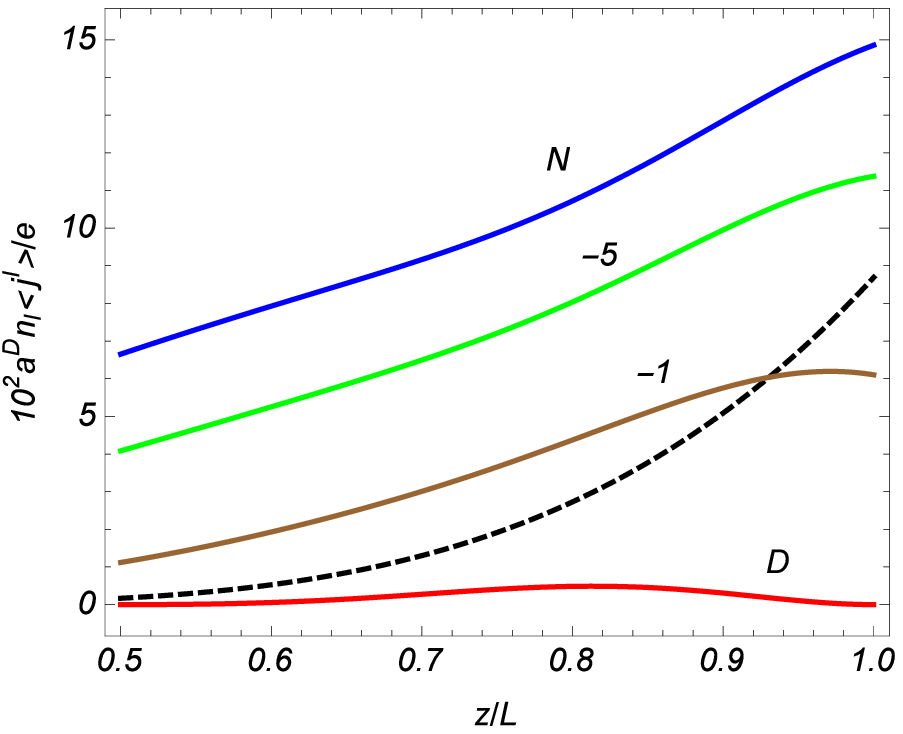,width=7.cm,height=5.5cm} & \quad %
\epsfig{figure=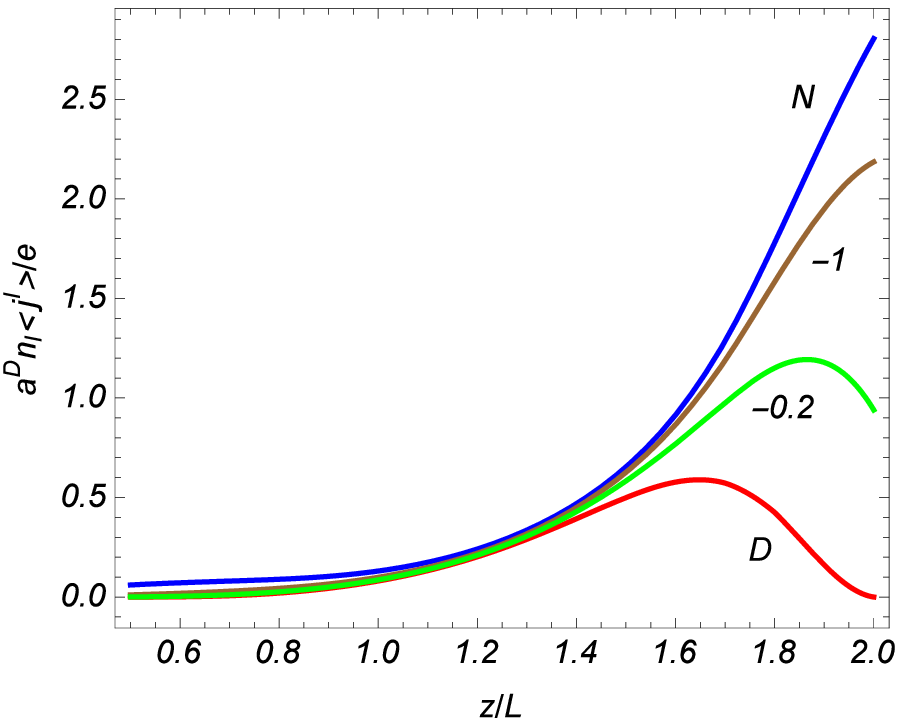,width=7.cm,height=5.5cm}%
\end{tabular}%
\end{center}
\caption{The current density as a function of $z/L$ for two different values
of the location of the right brane: $z_{2}/L=1$ (left panel) and $z_{2}/L=2$
(right panel). The graphs are plotted for Dirichlet, Neumann and Robin (the
numbers near the curves are the corresponding values of the ratio $\protect%
\beta /a$) boundary conditions for fixed $z_{1}/L=0.5$, $\tilde{\protect%
\alpha}=\protect\pi /2$.}
\label{fig5}
\end{figure}

In figure \ref{fig5}, the current density is plotted in the region between
the branes as a function of $z/L$ for $z_{2}/L=1$ (left panel) and $z_{2}/L=2
$ (right panel). For the other parameters we have taken $z_{1}/L=0.5$, $%
\tilde{\alpha}=\pi /2$. The graphs are presented for Dirichlet, Neumann and
for Robin boundary conditions (with the numbers near the curves being the
values of $\beta /a$). For $z_{2}/L=2$ the dependence of the roots of the
equation (\ref{ImmodeEq}) for imaginary modes on $\beta /a$ is qualitatively
similar to that depicted in figure \ref{fig2}. The dashed curve in the right
panel corresponds to the current density in the brane-free geometry.

\begin{figure}[tbph]
\begin{center}
\epsfig{figure=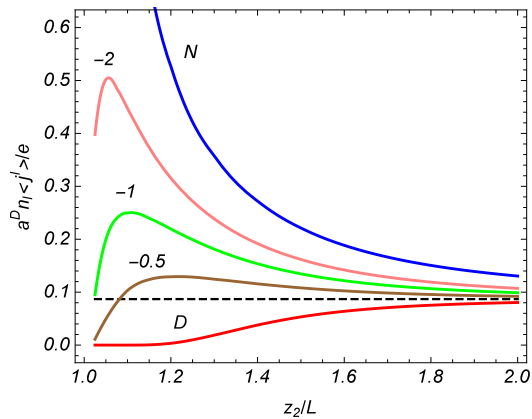,width=7.cm,height=5.5cm}
\end{center}
\caption{The current density at fixed observation point corresponding to $%
z/L=1$ as a function of the ratio $z_{2}/L$. The observation point has equal
proper distances from the left and right branes and, hence, $z_{1}/L=L/z_{2}$%
.}
\label{fig6}
\end{figure}

In order to show the dependence of the current density on the distance
between the branes, in figure \ref{fig6} the current density is displayed at
a fixed point $z/L=1$ as a function of the ratio $z_{2}/L$. The horizontal
dashed line corresponds to the current density in the absence of the branes.
We assume that the observation point has equal proper distances from the
left and right branes. This means that $z=\sqrt{z_{1}z_{2}}$ and, hence, for
the example in figure \ref{fig6} one has $z_{1}/L=L/z_{2}$. Under this
condition the proper distance between the branes is related to the ratio $%
z_{2}/L$ by $y_{2}-y_{1}=2a\ln (z_{2}/L)$. Hence, the figure \ref{fig6}
presents the current density as a function of the proper distance between
the branes at the fixed observation point in the middle between the branes.
For all the boundary conditions, except the Neumann one, the current density
tends to zero in the limit $y_{2}-y_{1}\rightarrow 0$. For Neumann boundary
condition the current density tends to infinity. This behavior of the VEV,
as a function of the interbrane distance, is in accordance with the general
analysis given above. For the values of the parameters we have taken one has
$\nu =D/2$ and in the case of Neumann boundary condition there is a special
mode with $\lambda =0$. The contribution diverging in the zero distance
limit comes from this mode. The contribution of the modes with $\lambda >0$
tends to 0 for the Neumann case as well. As it is seen from the graphs, in
the case of Dirichlet boundary condition, the current density (the modulus
of the current density for general $\tilde{\alpha}$) is smaller than the
boundary-free part and for the Neumann case it is bigger. In particular,
this means that the branes with Dirichlet (Neumann) boundary conditions
suppress (enhance) the vacuum currents.

\section{Applications in Randall-Sundrum type models with compact dimensions}

\label{sec:RS}

By using the results given above we can obtain the vacuum current densities
in generalized Randall--Sundrum braneworld models \cite{Rand99} with extra
compact dimensions. In these models the $y$-direction is compactified on an
orbifold, $S^{1}/Z_{2}$, with $-b\leqslant y\leqslant b$ and with the fixed
points $y=0$ and $y=b$. The latter are the locations of the hidden and
visible branes, respectively. The corresponding line element is given by (%
\ref{metric}) with the replacement $y\rightarrow |y|$. Because of this, the
Ricci scalar contains $\delta $-function terms located on the branes: $R=R_{%
\mathrm{AdS}}+4D[\delta (y)-\delta (y-b)]/a$. In addition, the action for a
scalar field may involve the contributions of the form%
\begin{equation}
S_{b}=-\frac{1}{2}\int d^{D+1}x\sqrt{|g|}[c_{1}\delta (y)+c_{2}\delta
(y-b)]\varphi ^{+}(x)\varphi (x),  \label{Sb}
\end{equation}%
where the constants $c_{1}$ and $c_{2}$ are the so-called brane mass terms.
Now, the equation for the radial part of the mode functions contains the $%
\delta $-function terms coming from the Ricci scalar and from the brane mass
terms. The boundary conditions for these functions are obtained by
integrating the equation near the branes. For fields even under the
reflection $y\rightarrow -y$ (untwisted scalar fields) the boundary
conditions obtained in this way are of the Robin type with the coefficients
(see \cite{Saha05,Gher00,Flac01})%
\begin{equation}
\frac{\beta _{j}}{a}=-\frac{2}{ac_{j}+4D\xi \delta _{j}},\;j=1,2.
\label{betjRS}
\end{equation}%
For odd fields (twisted scalars) Dirichlet boundary conditions are obtained
on both the branes.

Now the integration over $y$ in the normalization integral (\ref{Norm}) goes
over the range $-b\leqslant y\leqslant b$. As a consequence of this an
additional factor $1/2$ appears in the square of the normalization
coefficient of the mode functions. Hence, the expressions for the VEV of the
current density in the orbifolded braneworld models are obtained from those
given in the previous sections with an additional factor $1/2$ and with $%
y_{1}=0$, $y_{2}=b$.

\begin{figure}[tbph]
\begin{center}
\epsfig{figure=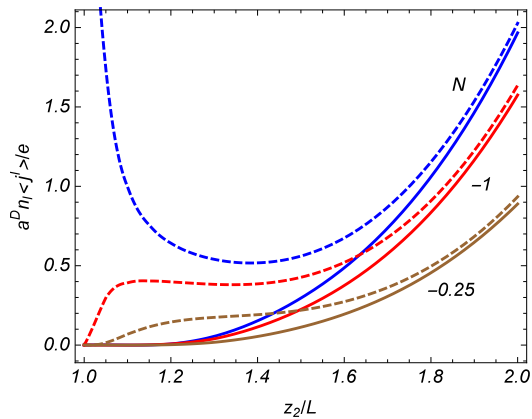,width=7.cm,height=5.5cm}
\end{center}
\caption{The current density on the visible brane for a minimally coupled
massless scalar field in the Randall--Sundrum model with a single extra
compact dimension as a function of the brane location. The graphs are
plotted for $L=a$ and $\tilde{\protect\alpha}=\protect\pi /2$ in the cases
of Dirichlet (full curves) and Neumann (dashed curves) boundary conditions
on the hidden brane. On the visible brane, Neumann and Robin (numbers near
the curves correspond to the values of $\protect\beta _{2}/a$) boundary
conditions are imposed. }
\label{fig7}
\end{figure}

In braneworld models of the Randall--Sundrum type the standard model fields
are located on the brane at $y=b$ (visible brane) and it is of interest to
consider the current density induced by a bulk scalar field on this brane.
In figure \ref{fig7}, in the model with $D=5$, $q=1$ (i.e. the
Randall--Sundrum model with a single compact extra dimension), the current
density is plotted on the visible brane, $z=z_{2}$, as a function of the
location of that brane for the length of the compact dimension $L=a$. For
the phase in the quasiperiodicity condition we have taken $\tilde{\alpha}%
=\pi /2$ and, as before, the graphs are plotted for a minimally coupled
massless scalar field. The full/dashed curves correspond to Dirichlet and
Neumann boundary conditions on the hidden brane. The numbers near the curves
correspond to the values of $\beta _{2}/a$ (the Robin coefficient for the
visible brane). In the numerical evaluations we have used the representation
(\ref{jlzj}) with $j=2$ and with an additional factor 1/2. Note that in the
case of Neumann boundary condition on both the branes the contribution of
the special mode with $\lambda =0$ should be added separately. This
contribution is given by (\ref{jlSp1}) with the function $\Omega ^{2}(z)$
from (\ref{OmN}) and, again, with an extra factor 1/2.

In the original Randall--Sundrum 2-brane model, the hierarchy problem
between the gravitational and electroweak scales is solved for the
interbrane distance about 37 times larger than the AdS curvature radius $a$.
In the setup we have considered this corresponds to large values of $z_{2}$
compared with $z_{1}$. If, in addition, one has $z_{2}\gg L_{i}$, the effect
of the hidden brane on the current density at the location of the visible
brane can be estimated by using the representation (\ref{jlonBr}) with $j=2$%
. The contribution induced by the hidden brane is given by the last term in
the right-hand side. The corresponding expression is simplified by taking
into account that for $z_{2}\gg L_{i}$ the argument $xz_{2}$ of the modified
Bessel functions is large in the integration range. The dominant
contribution comes from the region near the lower limit of the integral and,
under the assumption $|\tilde{\alpha}_{l}|\leqslant \pi $, from the term $%
\mathbf{n}_{q}=0$, corresponding to the lowest value of the momentum in the
compact subspace. By using the asymptotic expressions of the modified Bessel
functions for large arguments, to the leading order we find%
\begin{eqnarray}
\langle j^{l}\rangle _{\mathrm{RS},z=z_{2}} &\approx &\frac{1}{2}\langle
j^{l}\rangle _{0}+\frac{1}{2}\langle j^{l}\rangle _{1}^{(2)}-\frac{e\pi
^{(1-p)/2}z_{2}^{q+2}\tilde{\alpha}_{l}}{2^{p}a^{D+1}V_{q}}\,  \notag \\
&&\times \frac{\bar{I}_{\nu }^{(1)}(k_{(q)}^{(0)}a)}{\bar{K}_{\nu
}^{(1)}(k_{(q)}^{(0)}a)}(z_{2}k_{(q)}^{(0)})^{\frac{p-1}{2}%
}e^{-2k_{(q)}^{(0)}z_{2}},  \label{jlRs}
\end{eqnarray}%
where for the location of the hidden brane we have taken $%
z_{1}=ae^{y_{1}/a}=a$. The contributions $\langle j^{l}\rangle _{0}$ and $%
\langle j^{l}\rangle _{1}^{(2)}$ are given by the formulas from the previous
section with $z=z_{2}$. As is seen, the effects of the hidden brane on the
visible brane are suppressed by the exponential factor $\exp
(-2k_{(q)}^{(0)}z_{2})$.

\section{Summary}

\label{sec:Conc}

We have investigated the combined effects of the background geometry, the
nontrivial topology and the branes on the VEV of the current density for a
charged scalar field with a general curvature coupling parameter. In order
to have an exactly solvable problem, a highly symmetric locally AdS geometry
is considered with an arbitrary number of toroidally compactified spatial
dimensions. Along the compact dimensions the field obeys quasiperiodicity
conditions with arbitrary constant phases and, in addition, we have assumed
also the presence of a constant gauge field. By a gauge transformation, the
latter is equivalent to the shift of the phases in the periodicity
conditions equal to the magnetic flux enclosed by a compact dimension,
measured in units of the flux quantum. As the geometry of boundaries we have
considered two branes, parallel to the AdS boundary, on which the field
operator obeys Robin boundary conditions, in general, with different
coefficients. In the model at hand, all the properties of the vacuum state
can be extracted from the two-point functions and, as the first step, we
have evaluated the Hadamard function.

In the region between the branes the eigenvalues of the quantum number
corresponding to the coordinate perpendicular to the branes are roots of the
equation (\ref{Eigval}). In addition to an infinite number of modes with
positive $\lambda $, depending on the coefficients in the boundary
conditions on the branes, this equation may have modes with purely imaginary
$\lambda =i\eta $ and also modes with $\lambda =0$. In order to escape the
vacuum instability, we have assumed the condition (\ref{Stabil}) with $%
k_{(q)}^{(0)}$ being the minimal value of the momentum in the compact
subspace. Note that in the corresponding models with trivial topology the
presence of any mode with imaginary $\lambda $ leads to the vacuum
instability. The modes with $\lambda =0$ are present under the condition (%
\ref{Mode0Eq}) on the parameters of the model. For a given interbrane
distance, this condition gives a relation between the Robin coefficients. In
the expression of the Hadamard function, for the summation of the series
over the positive roots of the eigenvalue equation (\ref{Eigval}) we have
employed the generalized Abel-Plana formula. This allowed us to extract
explicitly the single brane contribution and to present the second
brane-induced part in terms of the integral rapidly convergent for points
away from the branes. Then, we have shown that the representations (\ref{G1}%
) and (\ref{G2}), obtained in this way, remain valid in the presence of the
modes with $\lambda ^{2}\leqslant 0$ if the condition (\ref{Stabil}) for the
stability of the vacuum is obeyed. Other representations, obtained by using
the summation formula (\ref{AbelPlan1}) for the series over the momentum
along the $l$th compact dimension, are given in Appendix. In these
representations the Hadamard function is decomposed into the contribution
corresponding to the model with an uncompactified $l$th dimension and the
part induced by the compactification of the latter to $S^{1}$. The first
term does not contribute to the component of the current density along the $l
$th dimension. The representations obtained in this way are well adapted for
the investigation of the VEVs on the branes.

Given the Hadamard function, the VEV of the current density is evaluated
with the help of the relation (\ref{jl1}). The VEVs of the charge density
and of the current components along uncompactified dimensions vanish. The
component of the current density along the $l$th compact dimension is
presented in two equivalent forms given by (\ref{jl2dec}) with $j=1,2$.
Another representation, in which the brane-induced contribution is presented
in the form of a single integral, is provided by (\ref{jl3}). The current
density along the $l$th compact dimension is an odd periodic function of the
phase $\tilde{\alpha}_{l}$ and an even periodic function of the remaining
phases $\tilde{\alpha}_{i}$, $i\neq l$. In both cases the period is equal to
$2\pi $. In particular, the current density is a periodic function of the
magnetic flux having the period equal to the flux quantum. In order to
clarify the behavior of the vacuum current as a function of the parameters
of the model, we have considered various limiting cases. First of all, we
have shown that in the limit of the large curvature radius of the background
spacetime the corresponding result for the Robin plates in the Minkowski
bulk with partially compactified dimensions is obtained. For a conformally
coupled massless field the current density in the AdS bulk is connected to
the one in the Minkowski spacetime by the conformal relation with an
appropriate transformation of the Robin coefficients (see (\ref{jlConf})).
In the limit when the brane at $z=z_{1}$ tends to the AdS boundary, $%
z_{1}\rightarrow 0$, the corresponding contribution to the current density
vanishes as $z_{1}^{2\nu }$ and, to the leading order, the result for the
geometry of a single brane at $z=z_{2}$ is obtained. For a fixed location of
the left brane, when the right brane tends to the AdS horizon, $%
z_{2}\rightarrow \infty $, its contribution to the vacuum current at a given
observation point decays as $\exp (-2k_{(q)}^{(0)}z_{2})$. If the length of
the $r$th compact dimension is much smaller than the remaining length scales
and the observation point is not too close to the branes, $L_{r}\ll |z-z_{j}|
$, for $0<|\tilde{\alpha}_{r}|<\pi $ the single brane contribution to the
current density is suppressed by the factor $\exp (-2|\tilde{\alpha}%
_{r}||z-z_{j}|/L_{r})$ for the brane at $z=z_{j}$, and the interference part
decays as $\exp [-2|\tilde{\alpha}_{r}|(z_{2}-z_{1})/L_{r}]$. In this limit,
the current density is localized near the branes within the region $%
|z-z_{j}|\lesssim L_{r}$. For small values of $L_{r}$ and $\tilde{\alpha}%
_{r}=0$, the component of the current density along the $r$th dimension
vanishes whereas the current density along other directions, to the leading
order, coincides with that in the $D$-dimensional model, with the excluded $r
$th compact dimension, divided by the proper length of the $r$th dimension.

For the investigation of the asymptotic for large values of the length $%
L_{l} $, it is more convenient to use the representation (\ref{jlAlt0}). In
this limit, the contribution to the VEV of the $l$th component of the
current density coming from the modes with positive $\lambda $ is suppressed
by the factor $\exp (-L_{l}\sqrt{\lambda _{1}^{2}+k_{(q-1)}^{(0)2}})$. If
the mode with $\lambda =0$ is present, for large values of $L_{l}$ the
corresponding current density decays as $\exp (-L_{l}k_{(q-1)}^{(0)})$ in
the case $k_{(q-1)}^{(0)}>0$ and as $1/L_{l}^{p+1}$ for $k_{(q-1)}^{(0)}=0$.
The representation (\ref{jlAlt0}) is also well adapted for the investigation
of the near-brane asymptotics of the current density. An important result
seen from this representation is the finiteness of the current density on
the branes. This feature is in drastic contrast compared to the cases of the
VEVs for the field squared and energy-momentum tensor. It is well known that
the latter diverge on the boundaries and in the evaluation of the related
global quantities, like the total vacuum energy, an additional
renormalization procedure is required. The current densities on the branes
are directly obtained from (\ref{jlAlt0}) by putting $z=z_{j}$ and are given
by the expression (\ref{jlzj}) with $j=1,2$ for the left and right branes,
respectively. In particular, for Dirichlet boundary condition the current
density and its normal derivative vanish on the branes. Another feature,
seen from the expression (\ref{jlAlt0}), is that the contribution to the
current density from the modes with positive $\lambda $ tends to zero in the
limit of small interbrane distances.

In the numerical examples, discussed in section \ref{sec:Curr}, we have
considered a minimally coupled massless scalar field in the $D=4$ model with
a single compact dimension and with the same Robin coefficients for the left
and right branes. In this case, for $\beta \leqslant 0$ there are no modes
with imaginary $\lambda $. In this range, for fixed values of the other
parameters, the current density is an increasing function of $|\beta |$. In
particular, it takes the minimum value for Dirichlet boundary condition and
the maximum value for the Neumann one. In the range $\beta >\beta _{c}>0$,
where $\beta _{c}$ is the critical value of the Robin coefficient for the
stability of the vacuum (the vacuum is unstable in the range $0<\beta <\beta
_{c}$), the situation is opposite: the current density decreases with
increasing $\beta $.

In section \ref{sec:RS} we have applied the general result to the 2-brane
Randall--Sundrum type model with extra compact dimensions. The corresponding
boundary conditions are obtained by the integration of the field equation
near the branes. For untwisted scalar fields the boundary conditions are of
the Robin type with the coefficients given by (\ref{betjRS}) with $c_{j}$
being the brane mass terms. For twisted fields Dirichlet boundary conditions
are obtained. The corresponding expressions for the vacuum currents are
obtained from those in section \ref{sec:Curr} with an additional coefficient
$1/2$ and taking $y_{1}=0$, $y_{2}=b$. For the values of the interbrane
distance solving the hierarchy problem between the electroweak and Planck
scales, the current density induced by the hidden brane on the visible one
is suppressed exponentially as a function of the location of the visible
brane.

\section*{Acknowledgments}

A. A. S. was supported by the State Committee of Science Ministry of
Education and Science RA, within the frame of Grant No. SCS 15T-1C110.

\appendix

\section{Other representations for the Hadamard function}

\label{sec:Appendix}

In section \ref{sec:Had}, by using the generalized Abel-Plana formula (\ref%
{SumAbel}), for the Hadamard function we have provided the representation (%
\ref{G1}). In the expression for the VEV\ of the current density obtained
from this representation (see (\ref{jl2dec})), the convergence of the
integral is too slow for points near the branes and it is not convenient for
the investigation of the near-brane asymptotic of the current density. Here
we apply to the mode sum for the Hadamard function another type of
Abel-Plana formula that allows one to extract the contribution induced by
the compactification. The representation obtained in this way is well suited
for the evaluation of the vacuum currents on the branes.

Let us apply to the series over $n_{l}$ in the representation (\ref{G}) the
Abel-Plana type summation formula \cite{Beze08}%
\begin{eqnarray}
&&\frac{2\pi }{L_{l}}\sum_{n_{l}=-\infty }^{\infty
}g(k_{l})f(|k_{l}|)=\int_{0}^{\infty }du[g(u)+g(-u)]f(u)  \notag \\
&&\qquad +i\int_{0}^{\infty }du\,[f(iu)-f(-iu)]\sum_{\lambda =\pm 1}\frac{%
g(i\lambda u)}{e^{uL_{l}+i\lambda \tilde{\alpha}_{l}}-1},  \label{AbelPlan1}
\end{eqnarray}%
where $k_{l}$ is given by (\ref{kl}). In the special case $g(u)=1$ and $%
\tilde{\alpha}_{l}=0$, this formula is reduced to the Abel-Plana formula in
its standard form. The contribution of the first term in the right-hand side
of (\ref{AbelPlan1}) gives the Hadamard function in the geometry of two
branes in $(D+1)$-dimensional locally AdS spacetime with compact dimensions $%
(x^{p+1},\ldots ,x^{l-1},x^{l+1},\ldots ,x^{D-1})$ and with the $l$th
dimension being uncompactified. The latter corresponds to the spatial
topology $R^{p+2}\times T^{q-1}$ and the respective Hadamard function will
be denoted by $G_{R^{p+2}\times T^{q-1}}(x,x^{\prime })$. As a consequence,
the Hadamard function is decomposed as%
\begin{equation}
G(x,x^{\prime })=G_{R^{p+2}\times T^{q-1}}(x,x^{\prime })+G_{l}(x,x^{\prime
}),  \label{Gdecl1}
\end{equation}%
where the last term comes from the second integral in the right-hand side of
(\ref{AbelPlan1}) and is induced by the compactification of the $l$th
dimension. It is presented in the form
\begin{eqnarray}
G_{l}(x,x^{\prime }) &=&\frac{a^{1-D}L_{l}(zz^{\prime })^{D/2}}{2^{p}\pi
^{p-1}V_{q}z_{1}}\sum_{s=1}^{\infty }\sum_{\mathbf{n}_{q-1}}\int d\mathbf{k}%
_{p}\,e^{ik_{r}\Delta x^{r}}\int_{0}^{\infty }dw\,\cosh (w\Delta t)  \notag
\\
&&\times \sum_{n=1}^{\infty }\lambda _{n}T_{\nu }(\chi ,\lambda _{n})g_{\nu
}(\lambda _{n}z_{1},\lambda _{n}z)g_{\nu }(\lambda _{n}z_{1},\lambda
_{n}z^{\prime })  \notag \\
&&\times \cosh \left( u\Delta x^{l}+is\tilde{\alpha}_{l}\right) \frac{%
e^{-suL_{l}}}{u}|_{u=\sqrt{w^{2}+\lambda _{n}^{2}+k^{(l)2}}},  \label{Gl2}
\end{eqnarray}%
where $\mathbf{n}_{q-1}=(n_{p+1},\ldots ,n_{l-1},n_{l+1},\ldots ,n_{D-1})$, $%
k^{(l)2}=k^{2}-k_{l}^{2}$, and the summation over $r$ in $k_{r}\Delta x^{r}$
is over $r=1,\ldots ,D-1$, $r\neq l$. Note that the part (\ref{Gl2}) is
finite in the coincidence limit of the arguments, including the points on
the branes. The physical reason for this feature is related to that the
toroidal compactification does not change the local geometry and the
structure of the divergences is the same as that for AdS bulk without the
compactification. The first term in the right-hand side (\ref{Gdecl1}) does
not contribute to the component of the current density along the $l$th
dimension.

Yet another representation is obtained applying to the series over $n$ in (%
\ref{Gl2}) the summation formula (\ref{SumAbel}). This leads to the
following decomposition%
\begin{eqnarray}
G_{l}(x,x^{\prime }) &=&G_{l}^{(1)}(x,x^{\prime })-\frac{8L_{l}(zz^{\prime
})^{D/2}}{(2\pi )^{p+2}a^{D-1}V_{q}}\sum_{s=1}^{\infty }\sum_{\mathbf{n}%
_{q-1}}\int d\mathbf{k}_{p}\,e^{ik_{r}\Delta x^{r}}  \notag \\
&&\times \int_{0}^{\infty }dw\,\cosh (w\Delta t)\int_{0}^{\infty
}dv\,\sum_{j=\pm 1}\cos \left( v\Delta x^{l}+js\tilde{\alpha}_{l}\right)
e^{-ijsvL_{l}}  \notag \\
&&\times \Omega _{1\nu }(uz_{1},uz_{2})X_{\nu }^{(1)}(uz_{1},uz)X_{\nu
}^{(1)}(uz_{1},uz^{\prime })|_{u=\sqrt{v^{2}+w^{2}+k^{(l)2}}}.  \label{Gl3}
\end{eqnarray}%
Here the term%
\begin{eqnarray}
G_{l}^{(1)}(x,x^{\prime }) &=&\frac{4\left( zz^{\prime }\right) ^{D/2}L_{l}}{%
\left( 2\pi \right) ^{p+1}a^{D-1}V_{q}}\sum_{n=1}^{\infty }\sum_{\mathbf{n}%
_{q-1}}\int d\mathbf{k}_{p}\,e^{ik_{r}\Delta x^{r}}\int_{0}^{\infty
}d\lambda \,\lambda  \notag \\
&&\times \frac{g_{\nu }(\lambda z_{1},\lambda z)g_{\nu }(\lambda
z_{1},\lambda z^{\prime })}{\bar{J}_{\nu }^{(1)2}(\lambda z_{1})+\bar{Y}%
_{\nu }^{(1)2}(\lambda z_{1})}\,\int_{0}^{\infty }dw\cosh (w\Delta t)  \notag
\\
&&\times \frac{e^{-nuL_{l}}}{u}\cosh (u\Delta x^{l}+in\tilde{\alpha}%
_{l})|_{u=\sqrt{w^{2}+\lambda ^{2}+k^{(l)2}}},  \label{Gl1br}
\end{eqnarray}%
comes from the first integral in the right-hand side of (\ref{SumAbel}). It
is the part of the Hadamard function induced by the compactification of the $%
l$th dimension in the geometry of a single brane at $y=y_{1}$.

\end{document}